\documentclass[traditabstract]{aa}

\usepackage{graphicx}
\usepackage{txfonts}

\usepackage{epsfig}
\usepackage{natbib}

\newcommand{\beq}{\begin{equation}}
\newcommand{\eeq}{\end{equation}}

\begin{document}

\title{Solar microflares: a case study on temperatures and the \ion{Fe}{xviii} emission}
\author{U.~Mitra-Kraev
\and
G.~Del Zanna 
}
\institute{DAMTP, Centre for Mathematical Sciences,  
 University of Cambridge,  Wilberforce Road, Cambridge CB3 0WA UK
}

 \date{Received 14 December 2018 / Accepted 16 May 2019}

 \abstract{In this paper, we discuss the temperature distribution and evolution of a microflare, simultaneously observed by Hinode XRT, EIS, and SDO AIA. 
 We find using EIS lines that during peak emission the distribution is nearly isothermal and peaked around 4.5~MK. 
This temperature is in good agreement with that obtained from the XRT filter ratio, validating the use of XRT to study these small events, invisible by full-Sun X-ray monitors such as GOES. 
 The increase in the estimated \ion{Fe}{xviii} emission in the AIA 94~\AA\ band can mostly be explained with the small temperature increase from the background temperatures. The presence of  \ion{Fe}{xviii} emission does not guarantee that temperatures of 7~MK are reached, as is often assumed. We also revisit with new atomic data the temperatures measured by a SoHO SUMER observation of an active region which produced microflares, also finding low temperatures (3--4~MK) from an \ion{Fe}{xviii}/\ion{Ca}{xiv} ratio.
 \keywords{Sun: corona -- Sun: flares --  Techniques: spectroscopic }
 }

\titlerunning{Solar microflares: a case study}

\maketitle

\section{Introduction }

The solution to the long-standing problem of solar coronal heating
remains elusive despite major advances in observational and
theoretical capabilities over the last few decades. 
Perhaps the most studied features of the solar corona are 
active regions (AR) which comprise a variety of structures, broadly classified as
warm loops [T $\simeq$ 1~MK], cool/fan loops [T $<$ 0.8~MK], 
and hot loops [T $\simeq $3~MK] which are in the cores of AR \citep[see, e.g., the reviews by ][]{reale:2012_lr,delzanna_mason:2018}. These hot core loops 
are nearly isothermal around 3~MK, 
and their characteristics do not change with AR evolution 
\citep[see, e.g.,][]{rosner_etal:78,saba_strong:1991,delzanna_mason:2003,delzanna:2013_multithermal,delzanna_mason:2014,delzanna_etal:2015_emslope}.

One of the theories  for the coronal heating assumes that 
nanoflare storms  occurring in the corona  heat the plasma
to temperatures higher than 3~MK, with subsequent cooling to form the 3 MK loops
\citep[see, e.g.,][]{parker:1988,klimchuk:2006,cargill:2014}.
Other theories, such as the dissipation of Alfv\'en waves
\citep[see, e.g.,][]{van_ballegooijen_etal:2011} also predict  high-temperature 
plasma heated by short-lived  events.

However, the presence of this hot plasma  has been a matter of much debate
in the literature, despite the availability of many observations, such as 
high-resolution extreme ultra-violet (EUV) images of the solar 
corona from the Solar Dynamics Observatory (SDO)  Atmospheric Imaging Assembly (AIA), 
 X-ray imaging  from the  Hinode X-Ray Telescope (XRT), as well as EUV spectra
from Hinode EUV Imaging Spectrometer (EIS). 

Many studies used a combination of the AIA, EIS and XRT instruments 
to find  significant hot emission above 3~MK. For example, 
\cite{warren_etal:2011,warren_etal:2012} combined AIA and EIS; 
\cite{winebarger_etal:2012} combined EIS with XRT,  while 
\cite{schmelz_etal:2015} combined AIA with XRT to improve 
the differential emission measure (DEM) at high temperatures. The limitations of such studies are due to the fact that 
the EIS lines formed above 3 MK are very weak and blended 
\citep[see, e.g.][]{delzanna:08_bflare,ko_etal:2008,delzanna_etal:2011_flare}, 
the AIA hot channels are strongly multi-thermal (in particular the 94~\AA\ band,
as summarised below), and  many of the XRT channels 
also have a broad temperature distribution. 
\cite{brosius_etal:2014} also found evidence for pervasive hot emission in \ion{Fe}{xix}, observed with the Extreme Ultraviolet Normal Incidence Spectrograph (EUNIS) sounding rocket flight in an AR core.

However, data  from EIS did not show any hot emission 
\citep{delzanna_etal:2011_aia,delzanna:2013_multithermal}
in quiescent  AR cores. A reanalysis of the brightest emission 
in active regions cores observed in the X-rays with the 
 Solar Maximum Mission (SMM)  X-ray polychromator (XRP) Flat  Crystal Spectrometer  (FCS)
 did not reveal any hot emission at least down to three orders of magnitude
from the peak at 3 MK \citep{delzanna_mason:2014}. 
Support to these results  have later been published  using observations 
from the Focusing Optics X-ray Solar Imager (FOXSI) sounding rocket payload
\citep[see, e.g.][]{ishikawa_etal:2014}, the  Nuclear Spectroscopic Telescope ARray (NuSTAR)
\citep[see, e.g.][]{hannah_etal:2016}, and more importantly using 
simultaneous  high-resolution spectra
from EIS and SoHO SUMER \citep{parenti_etal:2017}.
We refer the reader to this latter study for a more extended discussion on this topic.

One of the observational difficulties to study the hot core 3~MK loops is that they are 
unresolved with current instrumentation. Their unresolved nature was already known 
from ground-based coronal observations during eclipses or with 
coronagraph observations \citep[see, e.g.][]{ichimoto_etal:1995}.
We note that the emission at coronal temperatures between 
1.5 and 2.5 MK is also unresolved even at the highest spatial resolution
(0.25\arcsec) of the Hi-C instrument \citep[cf.][]{peter_etal:2013}, 
unlike the cooler loops which appear nearly resolved at 1\arcsec\ resolution
\citep{delzanna:2013_multithermal}.

Frequently, within the active region cores, small loop-like hotter structures appear. 
They appear resolved at 1\arcsec\ resolution, as routinely observed 
in the X-rays with XRT  and in the AIA 94~\AA\ band. 
The larger events are visible in the full-disk GOES X-ray monitor and range from A to  B-class. 
However, many smaller events are below the GOES threshold. 
We are interested to study these smaller events which we loosely term microflares.
SDO AIA 94~\AA\ images have been used by several authors 
\citep[see, e.g.,][]{warren_etal:2012} to show the common presence of 
 \ion{Fe}{xviii} emission, normally assumed in the literature to be formed around 7~MK, 
the temperature of the maximum emission of the ion in ionization equilibrium.
However, a combination of AIA and EIS observations  showed that 
some  \ion{Fe}{xviii} emission is often present even when the temperature 
of the plasma is about 3 MK  \citep{delzanna:2013_multithermal}.

This raises the question whether the microflares reach 7~MK at all, and how their 
temperatures can be measured.
Whether these events are frequent and ubiquitous enough to be able to produce the quiescent 3~MK core loops by cooling
 is another open question we would like to answer in a future study. 
We have selected several AR observations for this purpose, but in this paper we focus on 
 one particular event that  we present as a test case. 

In Sect.~\ref{overview} we present a short overview of previous microflare observations, 
and revisit temperatures obtained from an \ion{Fe}{xviii} observation of microflares by SUMER. In Sect.~\ref{data} we 
present our data selection and analysis procedures. We then discuss in some detail one microflare event observed simultaneously by 
 XRT, AIA and EIS (Sect.~\ref{case_study}), showing that the observations are all consistent with a maximum temperature of around 4--5~MK.
 We conclude with Sect.~\ref{conclusions}.

\section{A short overview  of microflare observations}\label{overview}

We start by noting that loop-like structures with 
lifetimes of about  10 minutes were often visible with the SoHO 
 Coronal Diagnostic Spectrometer (CDS) in 
\ion{Fe}{xix}, indicating temperatures above 5 MK. 
However, the CDS instrument, like EIS, is not ideal to study the evolution of such 
events, as the cadence of observations (when the slit
is `rastered' across an AR) is typically around 15 minutes. 
It is likely that these events, which are very common, are the same as the ones
observed by previous instruments, with lower or no spatial resolution. 
For example, increases in the X-ray signal with about 10-minute lifetimes
were commonly observed \citep{delzanna_mason:2014} 
with the Solar Maximum Mission (SMM) Bent Crystal Spectrometer (BCS), 
which had a field-of-view (FOV) of  about 6'$\times$6' (the size of a large active region).

The Yohkoh  Bragg Crystal Spectrometer (BCS) was used to 
measure temperatures in active regions, normally with the He-like Sulphur complex (with the isothermal assumption).
BCS typically measured a steady component of 3~MK, 
and a hotter, transient component around 5 MK or more, due to microflares
\citep[see, e.g.][]{watanabe_etal:1995}.
During quiescence, there was no indication of plasma with a temperature
above 3.5~MK. As the instrument observed the full Sun, 
perhaps the best measurements were made when only one AR was just behind the limb 
\citep{sterling_etal:1997}.

Interesting results were also obtained by \cite{feldman_etal:1996a} using the 
same Yohkoh BCS Sulphur complex. They 
studied 28 flares in the GOES X-ray class A2-A9, finding very low temperatures,
on average only 5~MK. 
The analysis and comparison with the GOES class was extended by 
\cite{feldman_etal:1996b}, where another interesting result was obtained:
a clear correlation between GOES class (i.e. energy content of the flare),
and the Yohkoh BCS temperatures during the peak of the X-ray emission. 
The data had a large scatter which was most likely due, according to the authors,
to the fact that the instrument was observing the full Sun, so there was 
always a bias created by all the active regions which were present on-disk
but were not flaring. 

The Yohkoh Soft X-ray Telescope (SXT) has also observed microflares in solar active regions \citep{Shimizu:1995}, with measured temperatures ranging between 4--8~MK, volume emission measure from $10^{44.5}$--$10^{47.5}~{\rm cm}^{-3}$ and lasting for 2--7~min.

Near-monochromatic images with the  CORONAS-F SPIRIT  
\ion{Mg}{xii} spectroheliograph showed recurrent brightenings in active regions
with typical timescales of 10 minutes \citep{reva_etal:2015}.
Unfortunately, \ion{Mg}{xii} has a broad temperature of formation in 
ionization equilibrium (between 5 and 25 MK), 
so it is not possible to assess if these are the same type of 
events as seen by  Yohkoh BCS.

The SphynX X-ray spectrometer aboard the CORONAS-Photon satellite 
unfortunately only operated over  a short period of time, during which one
AR crossed  the solar disk \citep[see, e.g.][]{sylwester_etal:2011}. 
The instrument had a sensitivity 100
times better than the GOES X-ray monitors, but the resolution and signal 
only allowed an isothermal analysis of the full-disk of the Sun. 
Temperatures between 2.5 and 6 MK were recorded during the AR passage. 
\cite{mrozek_etal:2018} studied in detail B-class flares observed by 
SphynX, while \cite{kirichenko_etal:2017} studied many weaker events, finding
interesting differences between the temperatures that would be estimated by extrapolating GOES and those actually estimated with SphynX.
However, one limitation of such observations is the lack of spatial resolution.

RHESSI observed a large number of small flares, always occurring 
in AR cores. For a review see \cite{hannah_etal:2008}.
Most of the events that have been studied with RHESSI are much larger than those considered here, though. 
Recently, NuSTAR, the very sensitive astrophysics mission, has also been used to observe quiescent active regions, and on one occasion a microflare was observed, as described in \cite{wright_etal:2017}. 
Although the NuSTAR resolution does not allow to resolve spectral lines, the (thermal) continuum measurements indicated very low temperatures (of the order of 3--5 MK) for the event of GOES class A0.1. Only during the impulsive phase a residual weak signal around 6.5~keV was measured, indicating that higher
temperatures, of the order of 10~MK, were possibly reached for a short period of time.

In summary, it is clear that full-Sun spectra and broad-band imaging 
have their limitations when studying microflares. 

\subsection{On microflares observed by SUMER in  \ion{Fe}{xviii}}

SoHO SUMER occasionally performed observations on active regions.
One interesting observation was described 
in \cite{teriaca_etal:2012} over a large active region on 2011 November 8 between 14:52 and 18:02 UT. During this time, several microflares occurred, so parts of them must have been recorded by the instrument while the slit was scanning the active region. 
\cite{teriaca_etal:2012} presented an analysis of the 
of the forbidden \ion{Fe}{xviii} 974~\AA\ and \ion{Ca}{xiv} 943~\AA\ lines. 

\begin{figure}[!htbp]
\centerline{\includegraphics[width=5.0cm,angle=90]{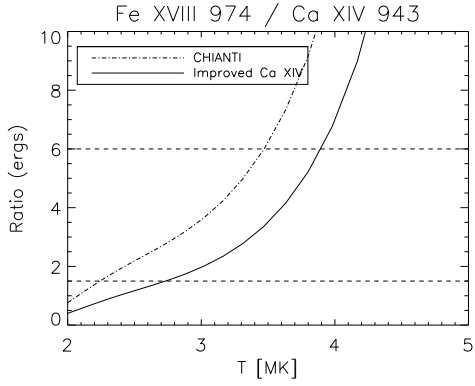}}
\caption{Theoretical curve of the ratio of  \ion{Fe}{xviii} vs.\  \ion{Ca}{xiv} 
observed by SUMER, using CHIANTI atomic data (dashed-dot line) 
and the new data (full line). The range of values observed by \cite{teriaca_etal:2012} in 
an active region core while several microflares were occurring are shown as dashed lines (i.e.\ within 1.5 and 6). 
The SUMER temperatures with the new atomic data are higher, but
still less than 4~MK. }
\label{fig:sumer}
\end{figure}
One puzzling result was that very low 
temperatures (around 2--3 MK) were obtained from the \ion{Fe}{xviii} vs.\  \ion{Ca}{xiv}
ratio. The two lines are not very far in wavelength so the relative calibration 
should be quite accurate. Both lines are relatively strong and unblended. 
It turns out that the low temperatures were partly caused by the atomic data
for  \ion{Ca}{xiv} as available within the CHIANTI database 
\citep{delzanna_etal:2015_chianti_v8}. 
We have used improved atomic data, which are briefly described 
in \cite{parenti_etal:2017}, to find increased temperatures, as shown in 
Fig.~\ref{fig:sumer}. 

Still, the maximum observed ratio values were around 6, indicating relatively low temperatures of 4~MK at most. 
It could well be that further improvements in the atomic data for 
\ion{Ca}{xiv} might increase the temperatures further, but not by much.
So, the direct spectroscopic information for one AR from SUMER 
confirms the \cite{delzanna:2013_multithermal} results, i.e.\ that 
the presence of  \ion{Fe}{xviii} emission in AR cores does not guarantee that temperatures of 7~MK are reached.

\section{Data selection and analysis} \label{data}

SDO AIA and Hinode XRT data were searched to identify suitable active regions. 
We focused on quiescent  ARs that were isolated, 
so that there was no interfering activity with
neighboring ARs, which  could cause heating events. ARs that exist from
limb to limb, or even better, over a full solar rotation, were preferred as they could
potentially be studied to see if their behaviour evolves over their full lifetimes.
GOES X-ray fluxes were analyzed to avoid flaring ARs. Various tools
such as Helioviewer were used.
As AIA observes the disk continuously, the main selection criterion 
was to find XRT multi-filter observations with at least 1~minute cadence, 
observing continuously for at least a few hours. 
We selected several ARs and microflares, and present here as a test case one event, as 
in this case we had also simultaneous Hinode EIS spectra with a high cadence (4~min),
which we could then use to validate the temperature results from XRT.

\subsection{Hinode - EIS}

We used custom-written software to process the EIS level-0 data
\citep[see, e.g.,][]{delzanna_etal:2011_aia}. We applied the \cite{delzanna:2013_eis_calib}
radiometric calibration. 
We fitted all the available line profiles with Gaussians, and co-aligned
spatially the two EIS channels with the AIA observations.

\subsection{Hinode - XRT}

We adopted the standard SolarSoft programs to process the data \citep{kobelski_etal:2014}. First, applying xrt\_prep with despiking and despotting (using keyword /despike\_despot) and co-aligning with AIA pointing as reference (keyword coalign=1), level 1 data is produced, along with grademaps which indicate bad (saturated, hot spots, etc.) areas/pixels. Then we removed jittering (xrt\_jitter) with reference to time 14:22.
XRT images were then co-aligned with respect to AIA, using the AIA 94~\AA\ \ion{Fe}{xviii} images. 

We have calculated the XRT temperature responses using the effective areas calculated for the specific dates, CHIANTI v.8 \citep{delzanna_etal:2015_chianti_v8}, and photospheric
abundances, although we note that the choice of chemical abundances does not 
affect the results based on the filter ratios presented here.
From the responses we obtained the isothermal temperatures using the filter ratios.
 \begin{figure}[tbp]
\centerline{\includegraphics[width=7.0cm,angle=0]{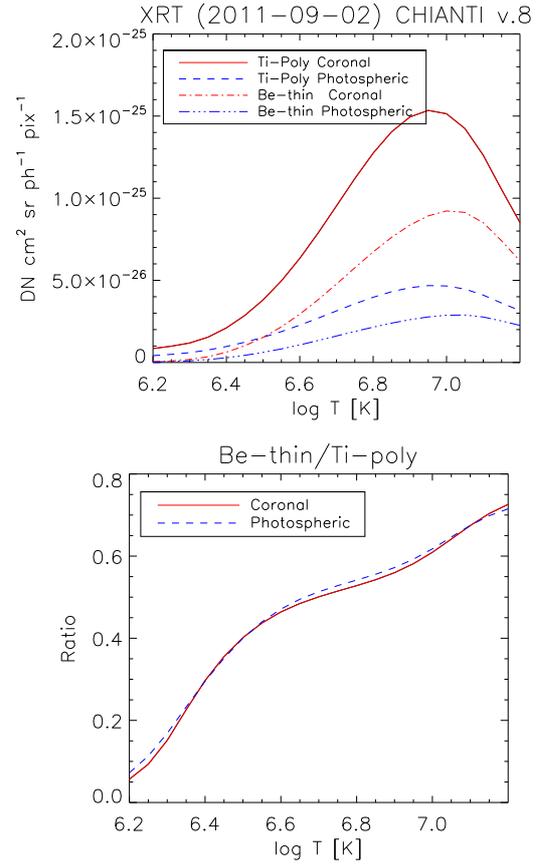}}
\caption{XRT Be-thin and Ti-poly temperature responses and their ratio.}
\label{fig:xrt_theo_ratio}
\end{figure}
Figure~\ref{fig:xrt_theo_ratio} shows one example, indicating that 
the ratio could in principle be used to measure temperatures between 
1 and 10~MK, although the signal towards lower temperatures decreases, so in effect
the filters are useful to measure temperatures above 2~MK.

\subsection{SDO - AIA}

We used the AIA cutout service, but applied our own cross-correlation 
to account for solar rotation, using the 1700~\AA\ images.
Despite having only about 1~\AA\ width, the AIA 94~\AA\ channel contains several 
emission lines, as discussed in \cite{delzanna:2013_multithermal}. This spectral band was 
chosen because of the presence of a hot \ion{Fe}{xviii} line at 93.932~\AA.
The images clearly showed the presence of multi-thermal material \citep{delzanna_etal:2011_aia}
and several studies were devoted to try and understand this band.

Significant progress was provided by \cite{delzanna:12_sxr1} with the identification of an 
 \ion{Fe}{xiv} 93.61\AA\ line, which is the dominant
contribution to the 94~\AA\ band in any AR observation, whenever
\ion{Fe}{xviii} is not present. Further progress was due to the improvement 
of the atomic data for the known \ion{Fe}{x} 94.012~\AA\ line 
with a new scattering calculation \citep{delzanna_etal:12_fe_10}.
The atomic calculations of \cite{odwyer_etal:12_fe_9} indicated instead that 
\ion{Fe}{viii} and \ion{Fe}{ix} contributions to the band are not large.

\cite{delzanna:2013_multithermal} used simultaneous EIS spectra and AIA 
observations, combined with the known ratios between the soft X-ray and EUV 
lines, to develop an atomic-based method to estimate the 
\ion{Fe}{xviii} count rates in the AIA 94~\AA\ channel,
using the 171 and 211~\AA\ as proxies for the 
\ion{Fe}{x} and \ion{Fe}{xiv} contributions. Namely, the count rate of the \ion{Fe}{xviii} emission is given by the count rates in the AIA 94~\AA\ band minus  the one in the AIA 211~\AA\ band divided by 120 minus the one in the AIA 171~\AA\  band divided by 450. We adopt this method here.
One result of the analysis was to indicate that significant 
\ion{Fe}{xviii} emission is often present in the cores of ARs
whenever a high EM at 3--4 MK was present.

\begin{figure}[tbp]
\centerline{\includegraphics[width=5.0cm,angle=90]{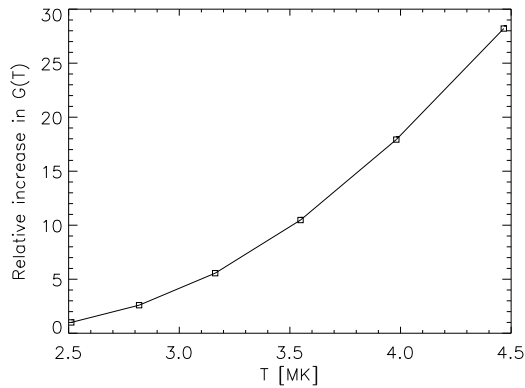}}
\caption{Relative increase in the $G(T)$ of the AIA \ion{Fe}{xviii}
line, normalised to 2.5 MK.}
\label{fig:fe_18_ratio}
\end{figure}

The contribution function $G(T)$ of the \ion{Fe}{xviii} line is peaked around 
7 MK, and has a steep increase from about 2.5 MK. 
We can simply estimate the increase in the AIA  \ion{Fe}{xviii} count rates
due solely to an increase in the  electron temperature by looking 
at the increase in the $G(T)$, normalized to the value at 2.5 MK,
as shown in Fig.~\ref{fig:fe_18_ratio}.
Even a small increase from 2.5 to 3.5 MK would result in an increase 
of the \ion{Fe}{xviii} count rates by a factor of 10.
Clearly, further increases can also be due to increased densities
(i.e. emission measures $EM$).

Finally, we note that other authors 
\citep[see, e.g.][]{warren_etal:2012} also developed various 
 methods to remove the cooler contributions to the 94~\AA\ band. 
However, all these other approaches were purely empirical.

\section{The microflare case study} 
\label{case_study}

On 2011-09-03, a series of recurrent microflares were observed 
by XRT and EIS within the small active region  NOAA 11283. 
The XRT instrument observed the AR with the Be-thin and Ti-poly
filters, with about 1~minute cadence. 
We have selected one of the main microflares.
 
 \begin{figure*}[!htbp]
    \centering
    \includegraphics[width=14.0cm,angle=0]{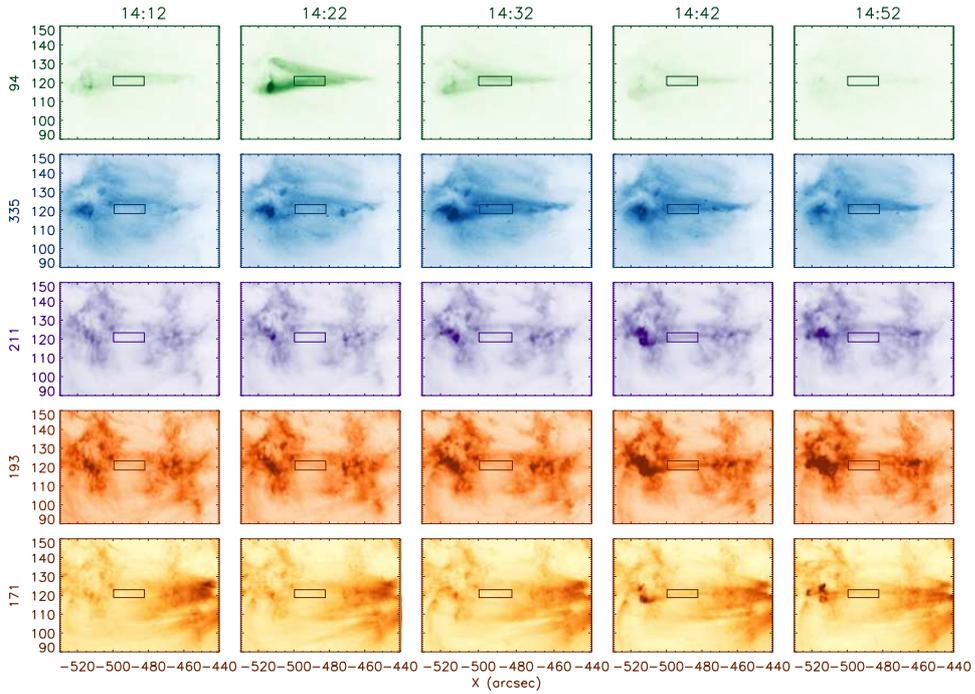}
    \caption{Sequence of AIA negative images in a selection of bands and timings, showing the 
evolution of the microflare which occurred on 2011-09-03 at 14:20 UT. 
The box indicates  the lower loop-top region chosen for a detailed study.}
    \label{fig:aia_images}
\end{figure*}
\begin{figure}[!htb]
    \centering
    \includegraphics[width=9cm]{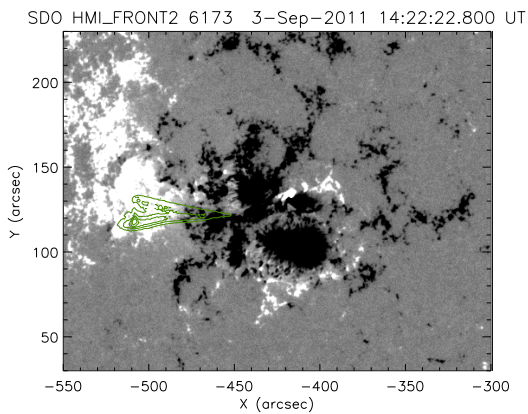}
    \caption{The SDO-HMI line-of-sight magnetogram of the active region NOAA 11283 at the time the microflare was observed, with the SDO-AIA 94\,\AA\ contour overlayed in green.}
    \label{fig:hmi}
\end{figure}

Figure~\ref{fig:aia_images} shows a sequence of AIA negative images 
in a selection of bands and timings, showing the evolution of the microflare, while Fig.~\ref{fig:hmi} displays the magnetic field configuration as observed by SDO-HMI with overlayed SDO-AIA 94\,\AA\ contours.
The event shows the appearance of 
two loops which become activated simultaneously. Some strong brightenings appear
at the footpoints of the loops, and indicate the locations of increased densities
in the regions where chromospheric evaporation is occurring 
\citep[for a discussion on signatures of chromospheric evaporation in small flares see][]{delzanna_etal:2011_flare}.
 Some other weak 
features appear to be connecting the footpoints of the loops and two nearby small 
jets which become activated at the same time (they are outside the FOV of the images). 
We selected a region near the top of the lower loop structure to 
obtain lightcurves and temperatures from XRT and EIS.

\begin{figure}[!htbp]
    \centering
    \includegraphics[width=9cm]{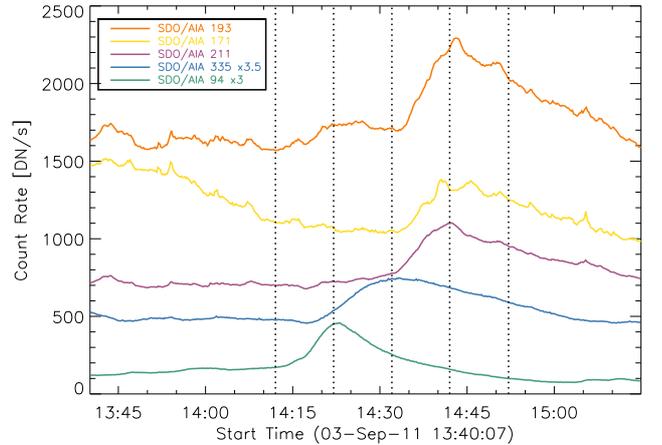}
    \caption{Scaled lightcurves in a selection of the  AIA bands, for the lower loop-top region
shown in  Fig.~\ref{fig:aia_images} (boxed area). The vertical dotted lines indicate the timings of the images in Fig.~\ref{fig:aia_images}.
}
    \label{fig:aia_lightcurves}
\end{figure}
Figure~\ref{fig:aia_lightcurves} shows the lightcurves in a selection of the  AIA bands. 
The peak emission in the AIA 94~\AA\ (\ion{Fe}{xviii}) is around 14:22. Within a few minutes, the loops
fade in \ion{Fe}{xviii} and reach peak emission in the 335~\AA\ band after 10 minutes, indicating
cooling to about 3 MK, as we interpret the peak emission in the 335~\AA\ band as being dominated
by \ion{Fe}{xvi}, which is is formed around 3~MK.
Interestingly, very little emission in the cooler AIA bands is present later on. 
Its spatial distribution is quite different from the one of the hotter \ion{Fe}{xviii}
and \ion{Fe}{xvi} emission. This behaviour is typical of these microflare events and is very different
than the behaviour of larger events, e.g. GOES  B-class \citep{delzanna_etal:2011_flare}
or C-class \citep{petkaki_etal:2012}, where the {\it same} loop-like structure as seen in 
higher temperatures is observed to cool down from about 10--12  MK through 5, 3, 1 MK, down to 
chromospheric temperatures. 

\begin{figure}[!htbp]
 \centerline{\includegraphics[width=6.0cm,angle=0]{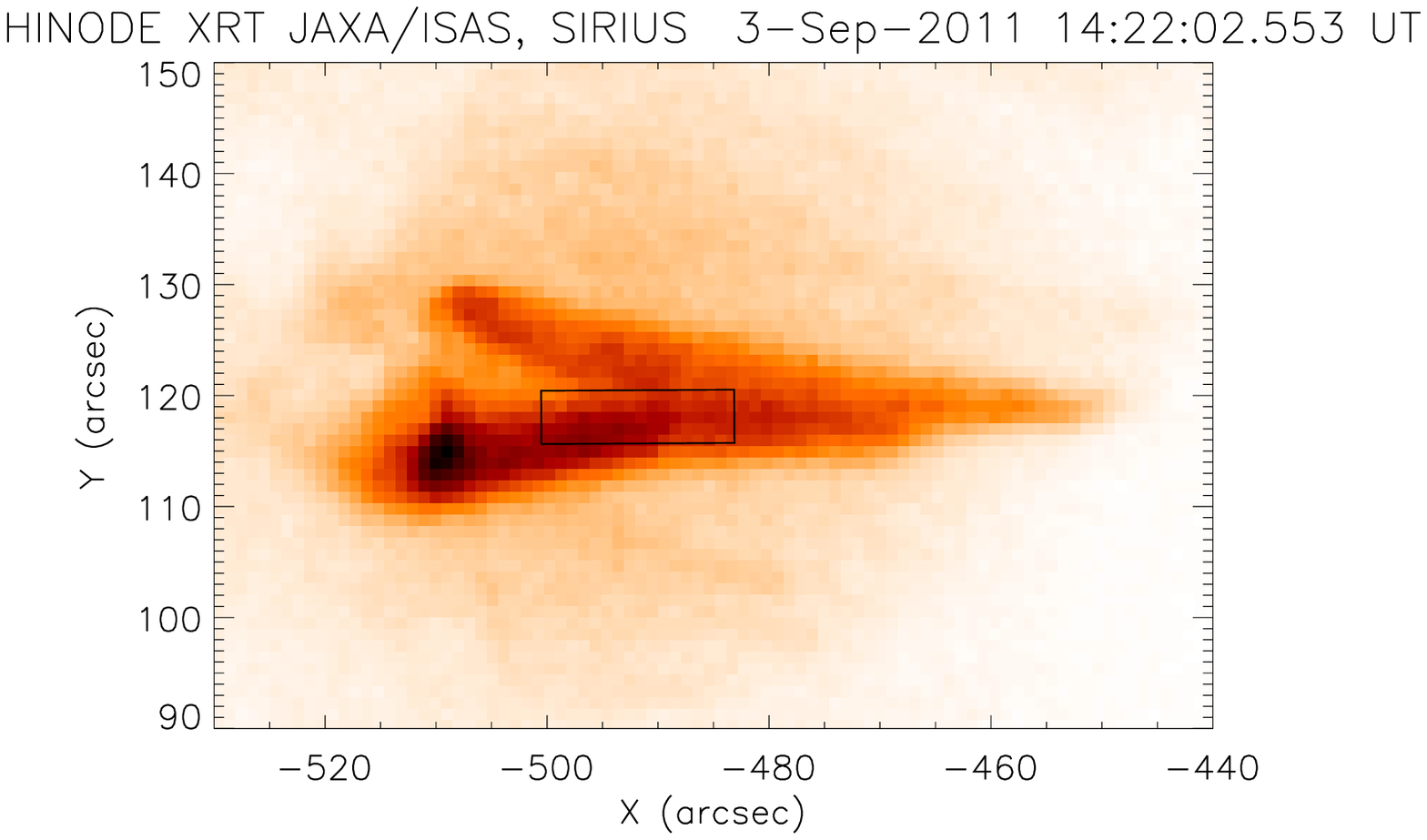}}
 \centerline{\includegraphics[width=6.0cm]{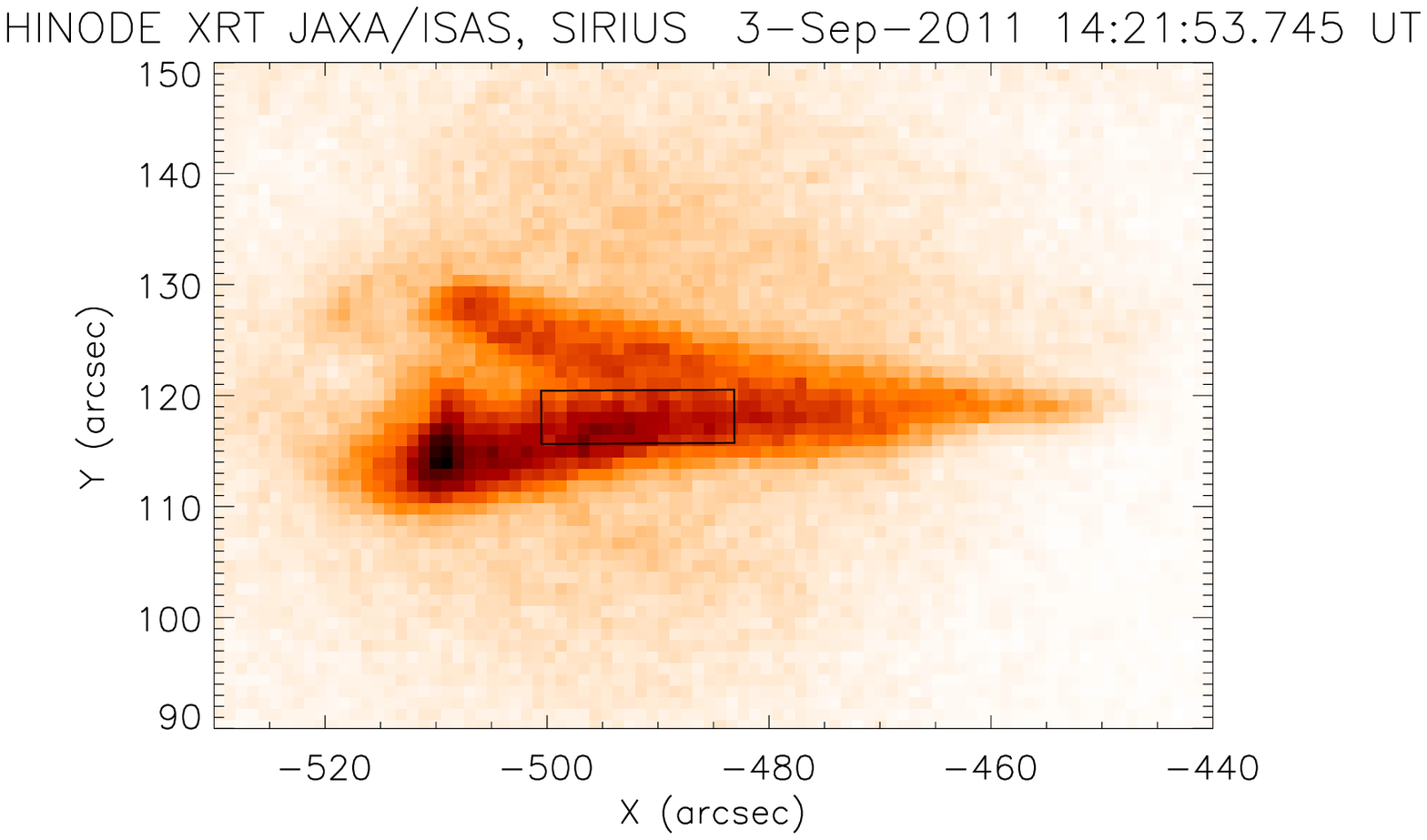}}
 \centerline{\includegraphics[width=6.0cm]{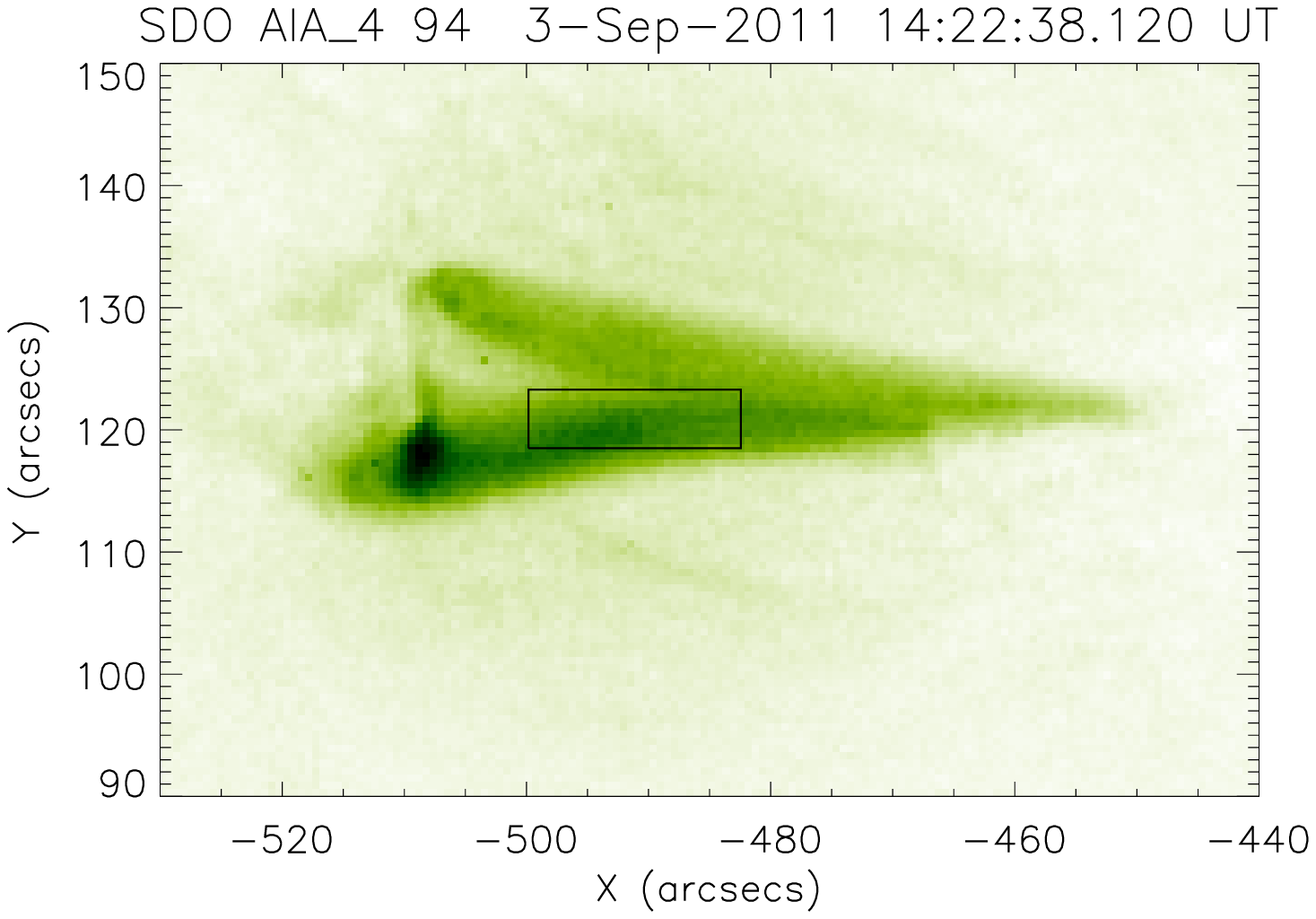}}
\caption{From top to bottom: Hinode XRT Ti-poly and Be-thin images during the peak of the microflare;
SDO AIA \ion{Fe}{xviii} emission. 
The box indicates the top of the lower loop area selected for further analyses. }
\label{fig:xrt_images}
\end{figure}
Figure~\ref{fig:xrt_images} shows a snapshot of the microflare
during the peak emission, as seen by two of the XRT bands (top two plots). As shown in the bottom plot of
Figure~\ref{fig:xrt_images}, the morphology of the microflare is nearly the same as seen in the XRT and
the  AIA \ion{Fe}{xviii} emission.

\begin{figure}[!htbp]
    \centering
    \includegraphics[width=8cm]{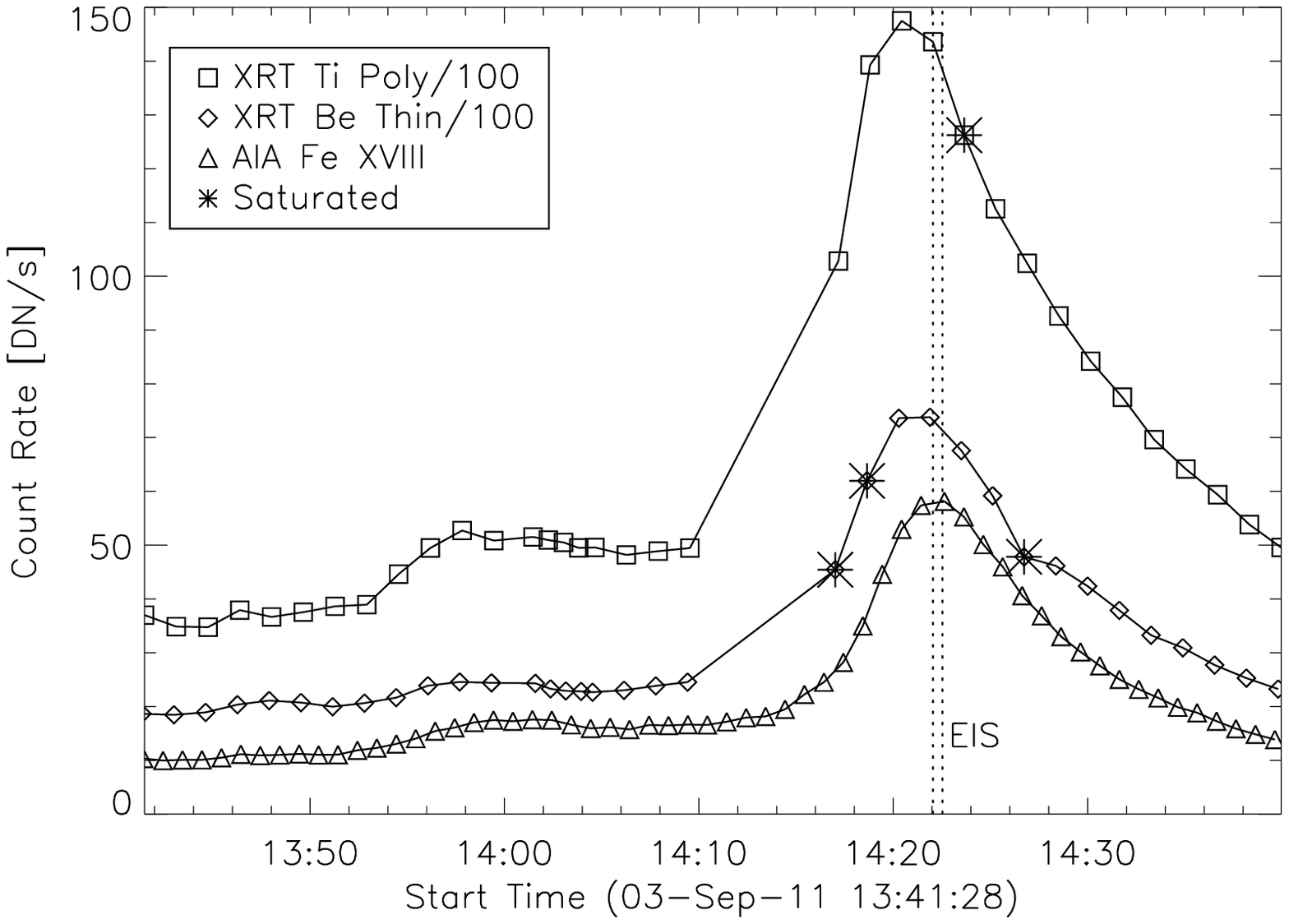}
 \centerline{\includegraphics[width=8.0cm]{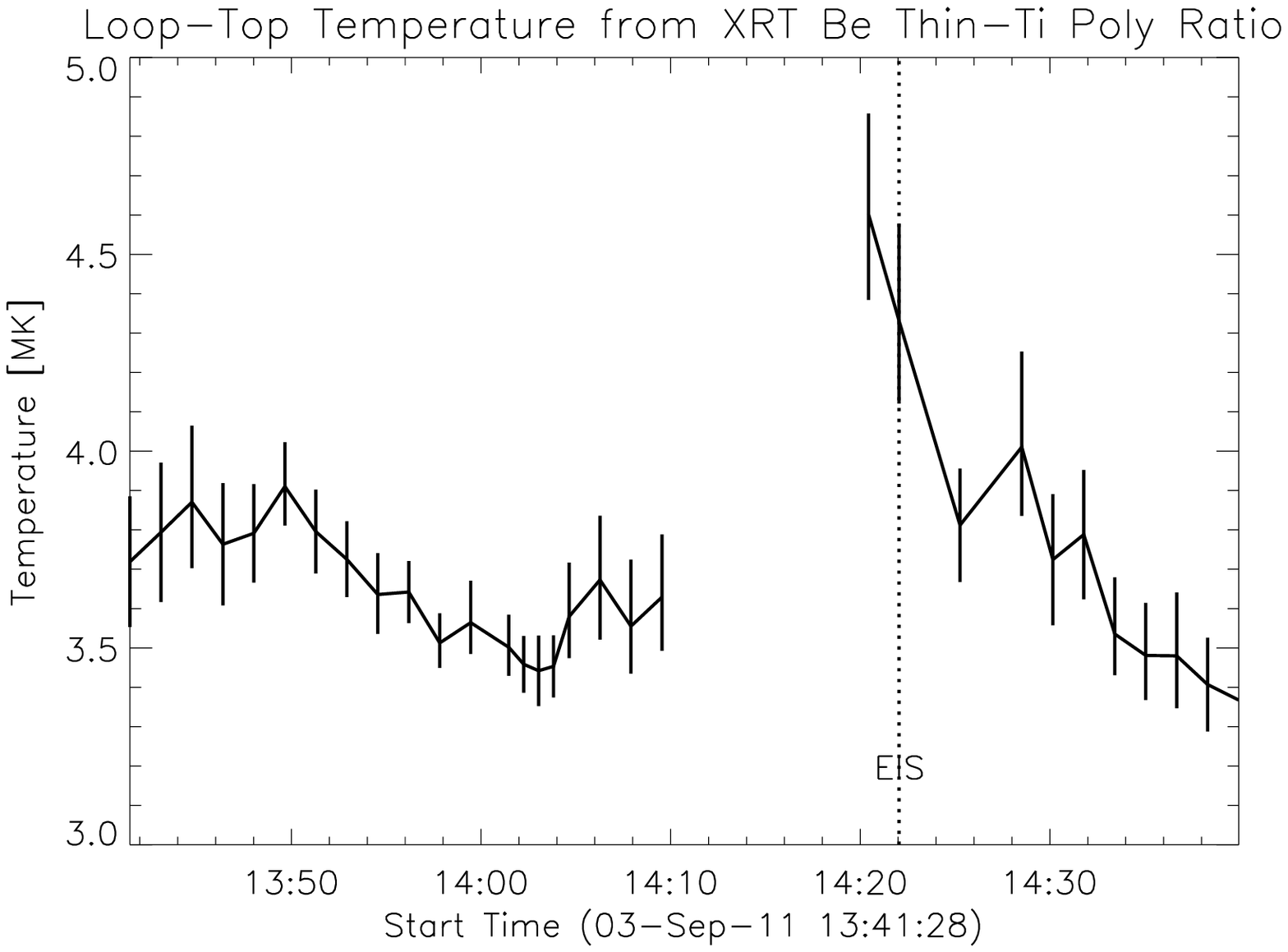}}
    \caption{Top: The averaged count rates of the data inside the box of the two XRT filters (Ti-Poly and Be-Thin, squares and diamonds respectively), both divided by 100, as well as the reconstructed Fe XVIII emission from the AIA filters (triangles). Saturated data points in XRT are marked with an asterix, and the vertical lines enclose the timing of the EIS observation of the same data interval.
Bottom: The temporal evolution of the temperature in the box, obtained from the XRT filter ratio.
 The dashed vertical line indicates the time of the EIS observation in the same region.}
    \label{fig:countrates}
\end{figure}
Figure~\ref{fig:countrates} (top) shows the lightcurves in the XRT bands  and AIA \ion{Fe}{xviii} in the loop top region.  
As a few of the XRT images were saturated (they are indicated with an asterix), we could not obtain 
XRT temperatures for those timings. 
The time differences in the two filters are within a few seconds, and we have checked that 
the intrinsic variations in the count rates do not affect the temperature measurements. 
The XRT temperatures for the loop top region for the timings not saturated are shown in 
the bottom plot of Fig.~\ref{fig:countrates}. 
The XRT indicates a temperature increase from about 3.5~MK to 4.5~MK. We have also obtained XRT temperatures
using the standard XRT SolarSoft programs, and obtained similar results. 
The uncertainties in the temperature reflect the Poisson noise in the XRT counts and 
take into account the theoretical variation of the ratio, as shown in Fig.~\ref{fig:xrt_theo_ratio}.

\begin{figure}[!htbp]
 \centerline{\includegraphics[width=7.0cm]{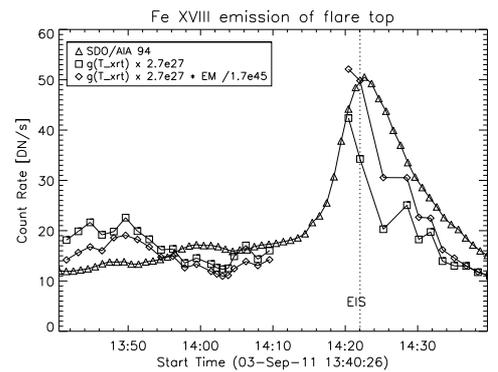}}
\caption{Relative increase in the AIA \ion{Fe}{xviii} signal in the lower loop top, as 
estimated from the  AIA  94~\AA\ signal (triangles), as obtained from the increased 
emissivity of the line due to the temperature variation obtained from XRT (squares),
and as obtained by combining the increase in the emissivity and the emission measure
(diamonds). The dashed vertical lines indicate the timings of the 
EIS observation during the microflare peak.
}
\label{fig:xrt_temp_increase}
\end{figure}

 The increase in the estimated  count rates due to  \ion{Fe}{xviii} in the AIA 94~\AA\ band in the loop-top region 
is about a factor of 5 during the peak. As we have measured the temperatures from XRT, it is interesting to 
see how much of the observed increase is due to the small 1 MK temperature increase (cf.  Fig.~\ref{fig:fe_18_ratio}).
We have therefore taken the XRT variation in temperature in the box and calculated the 
corresponding increase in the AIA  \ion{Fe}{xviii}  emission, shown in  
Fig.~\ref{fig:xrt_temp_increase}. This is about a factor of 4, indicating that the main increase 
in the AIA  94~\AA\ signal is due to such a small temperature increase.
The isothermal approximation used to measure temperatures from XRT also provide the values
of the emission measure, which are proportional to the square of the electron density. 
If we multiply the increase in the  \ion{Fe}{xviii} emissivity with the 
emission measure we obtain the points shown with diamonds in Fig.~\ref{fig:xrt_temp_increase},
which are quite close to the observed variation in the AIA  \ion{Fe}{xviii}  emission. 

Hinode EIS observed continuously the AR with the study
HH\_Flare+AR\_180x152, using the 
2\arcsec\ slit and 9s exposure times. The study was a `sparse raster'
of 30 slit positions with steps of 6\arcsec, covering a FOV of 
180x152\arcsec. Each slit position was observed every about 10s,
and the FOV was covered 
with a cadence of about 4 minutes. 
 Only a selection of a few spectral lines was telemetered to the ground.

 \begin{figure}
    \centering
    \includegraphics[width=9cm]{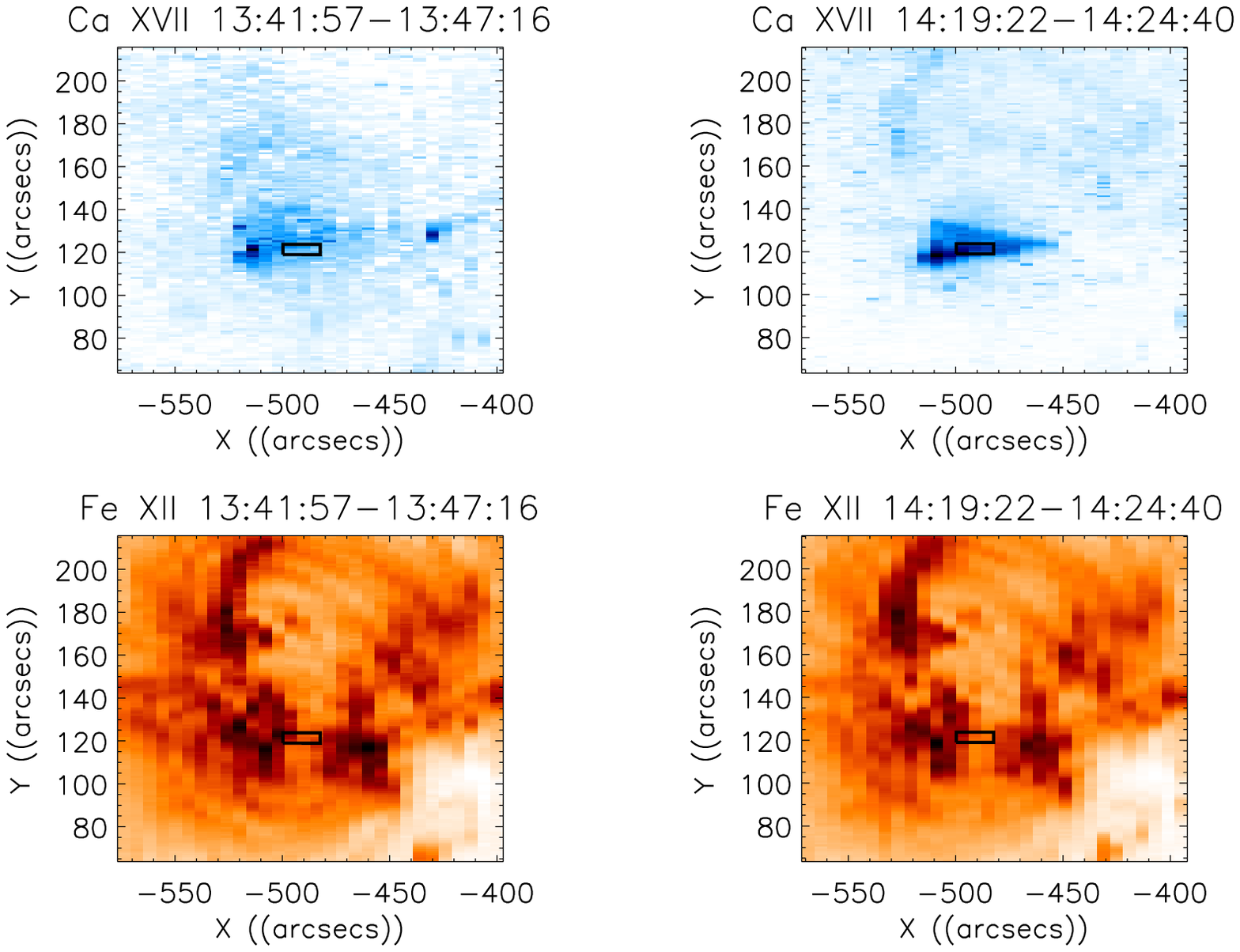}
    \caption{Hinode EIS monochromatic images in \ion{Fe}{xii} and \ion{Ca}{xvii} before
and during the main microflare. The box is drawn at the same location of the lower loop top as in the AIA and XRT images (Figs~\ref{fig:aia_images} \& \ref{fig:xrt_images}).}
    \label{fig:eis_images}
\end{figure}
\begin{figure}[!htbp]
 \centerline{\includegraphics[width=7.5cm]{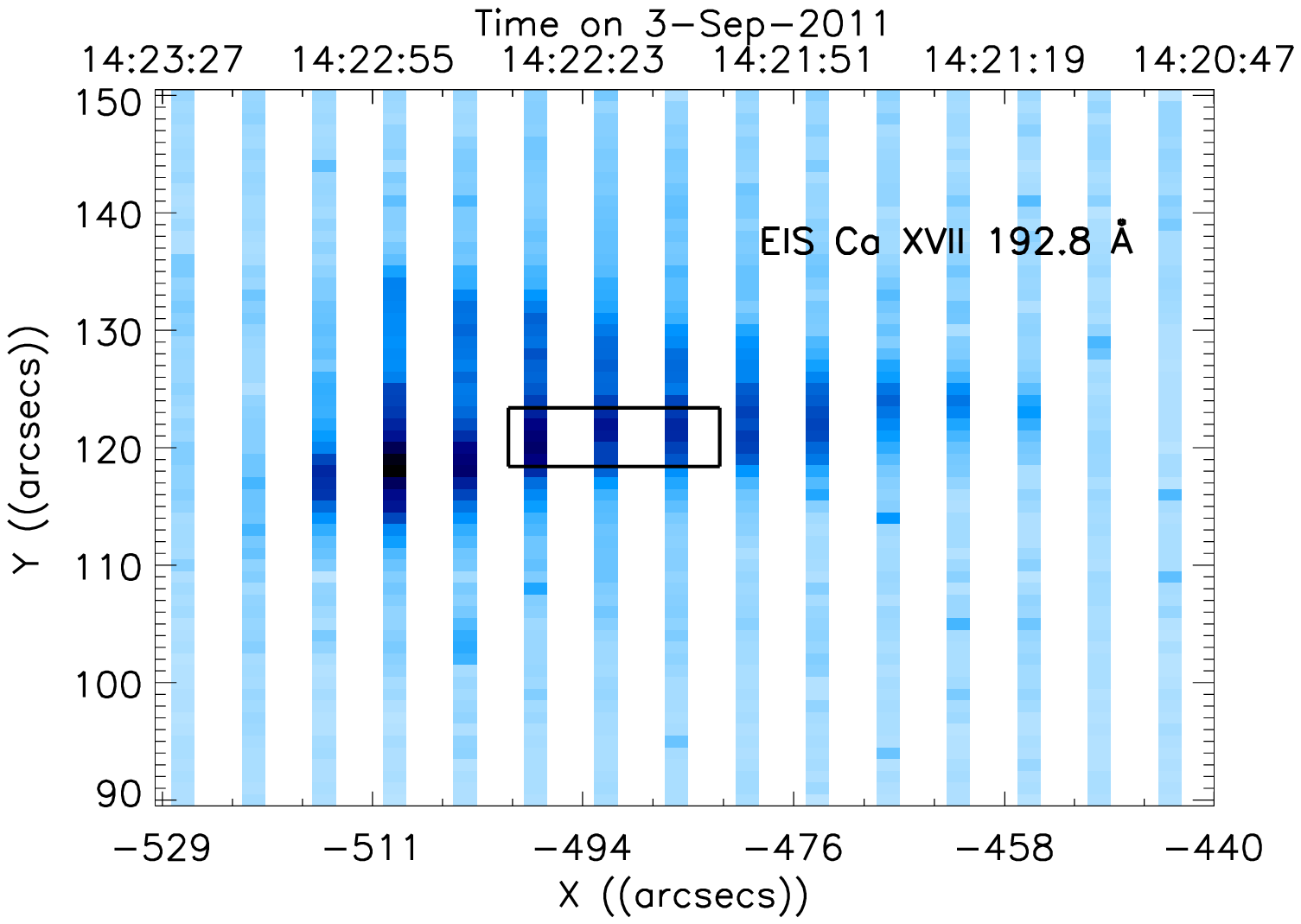}}
\caption{Showing the Hinode EIS monochromatic image of the \ion{Ca}{xvii} line at 192.8~\AA of the microflare. Due to the rastering, data were only recorded at sparse spatial resolution, the white stripes indicate data gaps. Each vertical raster is taken 11~s apart, the time-axis is given at the top of the graph, with time running from right to left.  The loop top is marked by the black box, which was selected for further analysis. The spatial coordinates of the data were shifted in order to be aligned with SDO AIA observations.}
\label{fig:eis_box}
\end{figure}

Figure~\ref{fig:eis_images} shows the monochromatic images obtained in the \ion{Fe}{xii} and \ion{Ca}{xvii} 
lines before (raster starting at 13:41 UT) and during (raster starting at 14:19 UT) the microflare.
Very little changes in the lines formed below 3~MK (as the \ion{Fe}{xii}) was observed. 
The hotter emission is clearly visible in \ion{Ca}{xvii}, and has the same morphology 
as the XRT and AIA \ion{Fe}{xviii}, as one would expect.
Brightenings associated with chromospheric evaporation are visible in the footpoint regions of the loops.
Figure~\ref{fig:eis_box} shows a zoomed in version of the \ion{Ca}{xvii} image during the peak of the microflare, 
indicating the starting times of each slit position and the location of the box in the lower loop top. 

Averaged EIS spectra in the box were obtained to increase the S/N ratio for the pre-flare and flare plasma.
All the lines were fitted and the calibration was later applied to obtain  
radiances. A sample of a few spectral windows is given in Appendix~\ref{add_microflare}.

\begin{figure}[!htbp]
 \centerline{\includegraphics[width=7.cm,angle=90]{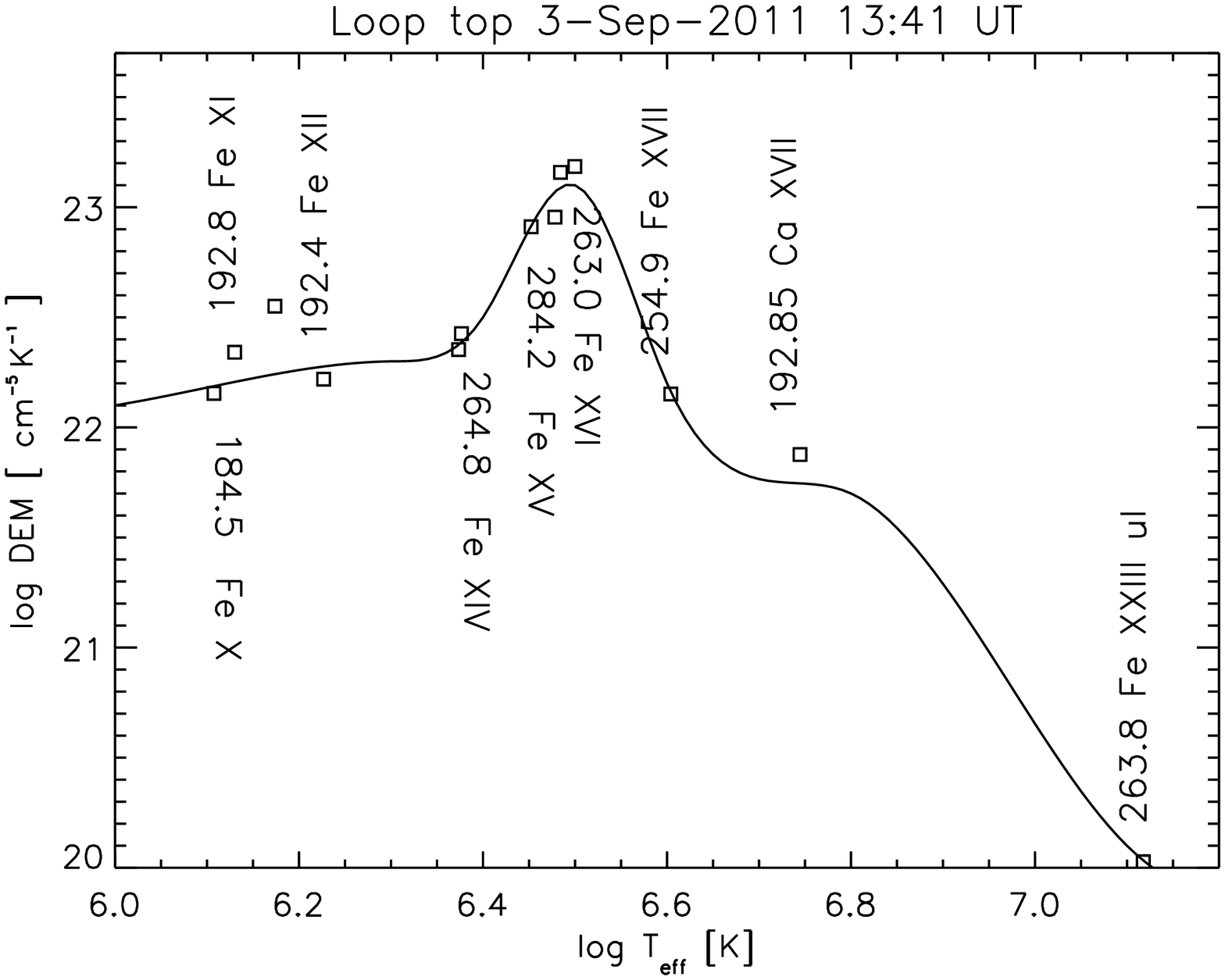}}
 \centerline{\includegraphics[width=7.0cm,angle=90]{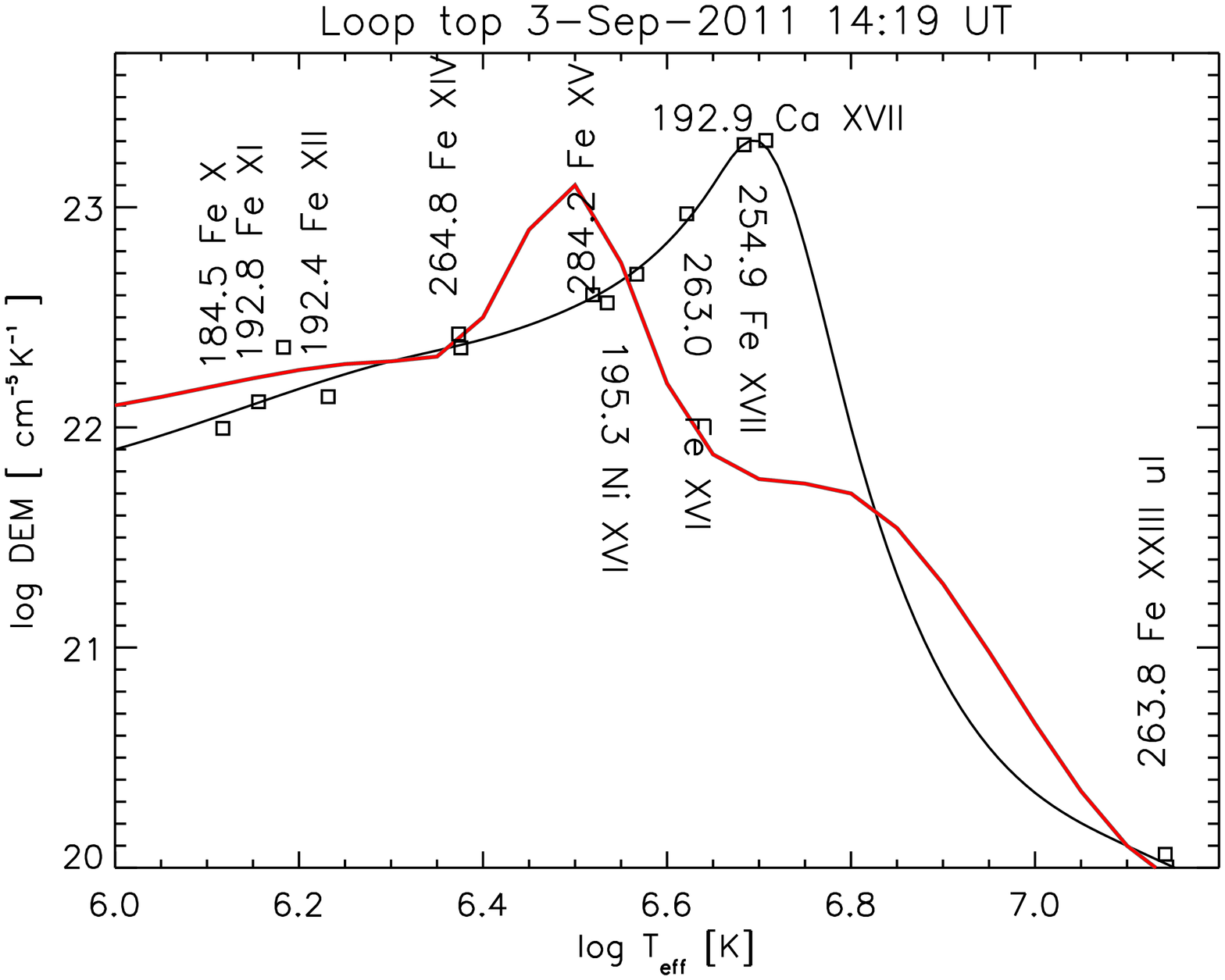}}
\caption{DEM before (top) and during the microflare (bottom), obtained from Hinode EIS. The pre-flare DEM is also shown in the bottom plot as a red line. 
The points are plotted at the effective temperature
$T_{\rm eff}$. The labels indicate the wavelength (\AA) and the main ion.}
\label{fig:eis_dem}
\end{figure}

To measure the temperature distribution (DEM) of the top of the lower loop, we assumed a spline functional form for the DEM and the method described in  \cite{delzanna_thesis99}.
The results for both  the  pre-flare and flare plasma 
are shown in  Fig.~\ref{fig:eis_dem}. 
The points are plotted at the effective temperature
$T_{\rm eff}$ (the formation $T$ of the line averaged over the DEM), and at the predicted vs. 
the observed intensity ratio, multiplied by the DEM value at $T_{\rm eff}$.
The respective values of the spectral lines used for the DEM fit shown in Fig.~\ref{fig:eis_dem} are given in Tables \ref{tab:flare} (flaring plasma) and \ref{tab:before_flare} (quiescent plasma), respectively.

\begin{table}[!htbp]
\caption{Observed  and predicted radiances for the microflare looptop region, raster start at 14:19 UT.}
\begin{center}
\footnotesize
\begin{tabular}{@{}llccrlrr@{}}
\hline\hline \noalign{\smallskip}
 $\lambda_{\rm obs}$  & $I_{\rm obs}$ & $T_{\rm max}$ & $T_{\rm eff}$  & $R$ & Ion & $\lambda_{\rm exp}$ &  $r$ \\
   (\AA)             &              &  (log)       & (log)         &     &     &  (\AA)   & \\
\noalign{\smallskip}\hline\noalign{\smallskip}
 192.93 & 81 &  5.42 &  5.56 &  0.64 &  \ion{O}{v} &  192.904 & 0.84 \\ 
 248.50 & 70 &  5.42 &  5.67 &  1.02 &  \ion{O}{v} &  248.461 & 0.78 \\ 
                                 &   &  &  &  &  \ion{Al}{viii} &  248.459 & 0.21 \\ 
 184.53 & 789 &  6.05 &  6.12 &  1.15 &  \ion{Fe}{x} &  184.537 & 0.93 \\ 
 192.80 & 415 &  5.42 &  6.16 &  0.99 &  \ion{Fe}{xi} &  192.813 & 0.92 \\ 
 192.38 & 905 &  6.20 &  6.23 &  1.20 &  \ion{Fe}{xii} &  192.394 & 0.97 \\ 
 264.79 & 2910 &  6.29 &  6.37 &  0.89 &  \ion{Fe}{xiv} &  264.789 & 0.94 \\ 
 274.21 & 2410 &  6.29 &  6.38 &  1.03 &  \ion{Fe}{xiv} &  274.204 & 0.95 \\ 
 192.02 & 130 &  7.19 &  6.50 &  1.03 &  \ion{Fe}{xxiv} &  192.028 & 0.14 \\ 
                                 &   &  &  &  &  \ion{Fe}{xi} &  192.021 & 0.76 \\ 
 284.16 & 30800 &  6.34 &  6.52 &  0.97 &  \ion{Fe}{xv} &  284.163 & 0.99 \\ 
 195.25 & 96.4 &  6.44 &  6.53 &  1.15 &  \ion{Ni}{xvi} &  195.271 & 0.91 \\ 
 262.99 & 4200 &  6.43 &  6.62 &  0.93 &  \ion{Fe}{xvi} &  262.976 & 0.99 \\ 
 254.90 & 463 &  6.74 &  6.68 &  1.01 &  \ion{Fe}{xvii} &  254.885 & 0.99 \\ 
 192.84 & 1720 &  6.76 &  6.71 &  0.94 &  \ion{Ca}{xvii} &  192.853 & 1.0 \\
 263.76 & 3 (ul) &  7.16 &  7.11 &  1.04 &  \ion{Fe}{xxiii} &  263.766 & 0.96 \\ 
\noalign{\smallskip}\hline 
\end{tabular}
\end{center}
\normalsize
 \label{tab:flare} 
\tablefoot{Each line lists the observed wavelengths 
 $\lambda_{\rm obs}$ (\AA), the measured  radiances $I_{\rm obs}$ 
(ergs cm$^{-2}$ s$^{-1}$ sr$^{-1}$), the maximum and effective temperature 
(log values, in K;  see text)  $T_{\rm max}$ and $T_{\rm eff}$,
 the ratio $R$ between the predicted and observed values, 
the main contributing ion and CHIANTI wavelength 
$\lambda_{\rm exp}$ (\AA), and the fractional contribution  $r$
(only contributions greater than 10\% are shown) to the blend. (ul) indicates an upper limit.
}
\end{table}

%
The `photospheric' abundances of \cite{asplund_etal:2009}
were adopted here as we would expect the flare plasma not to have had the time to develop coronal abundances, although we note that the DEM is well constrained by all iron lines, i.e., the chemical abundances only constrain the absolute value, not the shape.
The 3 MK peak in the pre-flare DEM is typical \citep{delzanna:2013_multithermal} 
and well constrained by lines from  \ion{Fe}{xv}, \ion{Fe}{xvi} and the hotter lines from \ion{Ca}{xvii} and \ion{Fe}{xvii}, for which upper limits were used. 

For the microflare, an upper limit to the 12~MK emission from 
\ion{Fe}{xxiii} 263.8~\AA\  was adopted.
 This was obtained by measuring the standard deviation of the 
counts where the line is, to obtain an estimate of the peak emission. 
This value was then converted to integrated radiances. The result is 3 ergs cm$^{-2}$ s$^{-1}$ sr$^{-1}$, i.e. the same obtained in \cite{parenti_etal:2017}.

The microflare peak DEM is well constrained by a significant increase in the \ion{Fe}{xvi} and especially the hot \ion{Ca}{xvii} and \ion{Fe}{xvii} emission. 
 The  \ion{Fe}{xvii} is weak but measurable, see Fig.~\ref{fig:spectra_microflare}.
As shown in Fig.~\ref{fig:spectra_microflare},
the \ion{Ca}{xvii} is significantly blended with \ion{Fe}{xi} and \ion{O}{v}
(see \cite{delzanna:08_bflare,delzanna_etal:2011_flare} for a discussion on blending for this line and the \ion{Fe}{xxiii} one). We fitted three Gaussians, with a careful selection of the limits in the wavelength and widths of these lines. 
The monochromatic image in \ion{Ca}{xvii} clearly shows that this method works relatively well when the line is strong. Normally, we would estimate the \ion{Ca}{xvii} by estimating the 
\ion{Fe}{xi} and \ion{O}{v}  contributions using other \ion{Fe}{xi} and \ion{O}{v} lines, but for the present study
this was not easily achievable as the other \ion{O}{v} line at 248.5~\AA\ was very weak,
and the other \ion{Fe}{xi} lines in the study were problematic. In fact, the \ion{Fe}{xi} 
192.63~\AA\ is known to be blended with an unidentified line, while the \ion{Fe}{xi} 192.~\AA\ is blended with at least \ion{Fe}{xxiv} and \ion{Fe}{viii} transitions. 

To cross-check that the \ion{Ca}{xvii} deblending was correct, we have added the \ion{Fe}{xi} and 
\ion{O}{v} deblended lines to the DEM analysis. The results in Table~\ref{tab:flare} clearly indicate an excellent agreement for the \ion{Fe}{xi} line, which is the main contributor to the blend. Furthermore, the fact that \ion{Ca}{xvii} is in excellent agreement with \ion{Fe}{xvii}
confirms that the deblending was successful. The \ion{O}{v} line at 248~\AA\ is a complex  blend and is very weak. This \ion{O}{v} line, compared to the \ion{O}{v} lines at 192.9~\AA\, is also density-sensitive and so it is difficult to estimate the \ion{O}{v} lines. 
However, it is clear that the \ion{O}{v} lines blended with \ion{Ca}{xvii} are not very strong, so even if their estimate is a factor of two wrong it would not significantly affect the \ion{Ca}{xvii} intensity.

The DEM obtained from the \ion{Ca}{xvii} and \ion{Fe}{xvii} EIS lines indicates a peak emission around 4.5~MK. Considering the pre-flare DEM, the lower loop emission is nearly isothermal. These results are important as they i) validate the isothermal assumption used to get temperatures from XRT and ii) confirm the XRT temperatures. 
Finally, we note that the difference in the DEM at 3~MK is somewhat misleading, as the Fe abundance 
at these temperatures is typically increased by a factor of 3.2 over its photospheric value \citep{delzanna:2013_multithermal}.

\begin{table}[!htbp]
\caption{Observed  and predicted (using the EIS DEM) Hinode XRT count rates}
\begin{center}
\begin{tabular}{@{}llccrlrr@{}}
\hline\hline \noalign{\smallskip}
Filter & Observed & Phot. & Phot.     &  Phot.    & Coronal \\
       &          &       & increased & decreased &        \\
\noalign{\smallskip}\hline\noalign{\smallskip}

Ti-poly &  14360 & 9328 &  9749 & 9303 &  7006  \\
Be-thin &   7380 & 4589 &  4897 & 4589 &   3420 \\
\noalign{\smallskip}\hline 
\end{tabular}
\end{center}
\normalsize
 \label{tab:xrt_cr} 
\tablefoot{The Phot.\ values are obtained with photospheric 
abundances, the last column shows the values with coronal abundances,
where Fe is increased by a factor of about 4 \citep{feldman_etal:1992}.
The `increased' (`decreased') values are those obtained when the DEM at log $T$=7.1
is increased (decreased) by a factor of 10. }
\end{table}

\begin{table}[!htbp]
\caption{Observed and predicted radiances for the looptop region before the microflare 
(same layout as Table~\ref{tab:flare}), raster start at 13:41 UT.}
\begin{center}
\footnotesize
\begin{tabular}{@{}llccrlrr@{}}
\hline\hline \noalign{\smallskip}
 $\lambda_{\rm obs}$  & $I_{\rm obs}$ & $T_{\rm max}$ & $T_{\rm eff}$  & $R$ & Ion & $\lambda_{\rm exp}$ &  $r$ \\
   (\AA)             &              &  (log)       & (log)         &     &     &  (\AA)   & \\
\noalign{\smallskip}\hline\noalign{\smallskip}

 192.93 & 120 &  5.42 &  5.52 &  1.07 &  \ion{O}{v} &  192.904 & 0.86 \\ 
 
 248.50 & 210 &  5.42 &  5.61 &  0.80 &  \ion{O}{v} &  248.461 & 0.84 \\ 
                                 &   &  &  &  &  \ion{Al}{viii} &  248.459 & 0.14 \\ 
 
 184.53 & 1200 &  6.05 &  6.11 &  1.08 &  \ion{Fe}{x} &  184.537 & 0.93 \\ 
 
 192.80 & 756 &  5.42 &  6.13 &  0.73 &  \ion{Fe}{xi} &  192.813 & 0.88 \\ 
 
 
 192.38 & 1130 &  6.20 &  6.23 &  1.14 &  \ion{Fe}{xii} &  192.394 & 0.97 \\ 
 
 274.21 & 3030 &  6.29 &  6.37 &  1.05 &  \ion{Fe}{xiv} &  274.204 & 0.95 \\ 
 
 264.79 & 3610 &  6.29 &  6.38 &  0.91 &  \ion{Fe}{xiv} &  264.789 & 0.95 \\ 
 
 284.16 & 35000 &  6.34 &  6.45 &  1.01 &  \ion{Fe}{xv} &  284.163 & 0.99 \\ 
 
 195.25 & 130 &  6.44 &  6.48 &  1.30 &  \ion{Ni}{xvi} &  195.271 & 0.91 \\ 
 
 192.02 & 206 &  7.25 &  6.48 &  0.85 &  \ion{Fe}{xxiv} &  192.028 & 0.12 \\ 
                                 &   &  &  &  &  \ion{Fe}{xi} &  192.021 & 0.74 \\ 
                                 &   &  &  &  &  \ion{Fe}{viii} &  192.004 & 0.11 \\ 
 
 262.99 & 3300 &  6.43 &  6.50 &  0.82 &  \ion{Fe}{xvi} &  262.976 & 0.98 \\ 
 
 254.90 & 108 &  6.74 &  6.60 &  1.03 &  \ion{Fe}{xvii} &  254.885 & 0.99 \\ 
 
 192.84 & 213 &  6.76 &  6.74 &  0.74 &  \ion{Ca}{xvii} &  192.853 & 1.0 \\
 
 263.76 & 3 (ul) &  7.16 &  7.12 &  1.01 &  \ion{Fe}{xxiii} &  263.766 & 0.96 \\ 

\noalign{\smallskip}\hline 
\end{tabular}
\end{center}
\normalsize
 \label{tab:before_flare} 
\end{table}

Having obtained the DEM of the loop top, we can then easily predict the signal in the XRT bands and the estimated AIA \ion{Fe}{xviii}  
using CHIANTI and the effective areas of the instruments for the specific date as available within SolarSoft.
For the forward modelling, we used the same photospheric abundances as the ones which were used for the DEM inversion, and note that some differences would be present in the predicted XRT count rates if other abundances were used. 
Table~\ref{tab:xrt_cr} lists these predicted count rates versus the observed ones for both the XRT Ti-poly and the Be-thin filters, for both photosperic as well as coronal abundances. 
We have calculated the XRT count rates adding the continuum and summing all contributions from 2 to 200~\AA, using CHIANTI data. 
The observed count rates in the XRT bands are about 50\%  higher than predicted. 
We note that the ratios of the two bands are nearly the same (about 2) for all the different columns, observed as well as predicted, regardless of the abundance model, which is reassuring and gives us confidence in the reliability of using this filter ratio to derive a temperature, which in this case is around 4.5~MK (see also Fig.~\ref{fig:xrt_theo_ratio}).

The EIS DEM was obtained with a spline node at log $T$=6.7, one at 6.8 and the last one 
at 7.1. The last point was kept fixed. The DEM at log $T$=6.8 cannot 
be increased as this would substantially increase the \ion{Ca}{xvii} and \ion{Fe}{xvii} line fluxes, which are very sensitive to this temperature. The DEM at log $T$=7.1 is unconstrained as we do not have a measurement of the \ion{Fe}{xxiii} line. We have calculated the XRT count rates obtained by increasing and decreasing the DEM at  log $T$=7.1  by a factor of 10, and added these results to Table~\ref{tab:xrt_cr}. We can see that the two XRT filters are not very sensitive to these changes. This can be explained by the fact that most of the XRT count rates in these two filters come from \ion{Fe}{xvii} lines and other lines formed at similar temperatures, as seen in Fig.~\ref{fig:xrt_sp_microflare} and previously discussed in  \cite{odwyer_etal:2014}. We also note that increasing the DEM above 10~MK would increase the emission in the \ion{Fe}{xxiv} 192.0~\AA\ line, which would be inconsistent with observation. 
 
 We have also tested the effects on the predicted XRT counts related to the chemical abundances. We have run the 
DEM fitting by adopting the `coronal' abundances of \cite{feldman_etal:1992}, where the iron abundance is increased by a factor of about 4. We have then run the forward modelling, and obtained the XRT count rates in the last column of Table~\ref{tab:xrt_cr}. We can see that assuming coronal abundances worsens the discrepancy between the observed and predicted XRT count rates (from a factor of 1.5 for photospheric to a factor of 2 for coronal abundances).

In the case of AIA, the EIS DEM predicts 94 DN/s/pixel for the  \ion{Fe}{xviii} emission\footnote{If, additionally to using the date-specific effective area files available within SolarSoft, calibration corrections accounting for differences between AIA and EVE (using SolarSoft keyword /eve\_norm) are made, the predicted count rate for the \ion{Fe}{xviii} emission  is even higher (117~DN/s/pixel).}. The lower-temperature Fe emission in the AIA band is predicted to be much lower: 2.3 DN/s/pixel for \ion{Fe}{xiv} and 1.3 for  \ion{Fe}{x}. The observed count rates are about 60, i.e. about 50\% lower than predicted. This difference could also be due to calibration problems, although we note
that the  high-temperature tail in the DEM is not well constrained by the EIS observations, so a difference of 50\% is not very large.

\subsection{Radiometric calibration issues}

We suggest that the observed discrepancies are due to 
calibration issues with the XRT and the AIA 94~\AA\ band. This issue is complex and cannot be easily resolved. We note here a few key points and refer to the Appendix~\ref{2011Oct22} for further information. 

The fact that the XRT observed count rates are higher than expected was reported by several authors. 
\cite{wright_etal:2017} also found a difference of a factor of two when comparing the well-calibrated NuSTAR observations of a microflare with XRT; on the other hand, \ion{Fe}{xviii} flux obtained from the AIA bands was in excellent agreement with NuSTAR. These observations were in 2015. They adopted the `coronal' abundances of \cite{feldman_etal:1992}. With the same coronal abundances we also find a factor of 2 discrepancy with XRT.

Earlier, \cite{schmelz_etal:2015} suggested a discrepancy of a factor of two in the absolute values of the XRT effective area, for observations in 2011, using coronal abundances \citep{schmelz_etal:2012}.

We note that relatively good agreement ($\sim$30\%) was
found by \cite{odwyer_etal:2014} cross-correlating EIS and XRT observations from Dec 2007, using coronal abundances, while larger descrepancies ($\sim$50\%) were found using photospheric abundances.

Original calibration of XRT in the first 2.5 years of operation \citep{narukage_etal:2011} for the quiet Sun show that the counts in all the XRT filters have significant decrease in sensitivity (some by a factor of 3 within 5 months). Corrections were implemented. Frequent bakeouts were introduced to increase the sensitivity. A follow-up paper by \cite{narukage_etal:2014} adopted the relative calibration of Ti-poly and Be-thin filters to constrain the calibration of the thick filters, imposing that the same temperatures would be obtained from the various filter combinations.

As far as we are aware, there are no published studies of the long-term absolute calibration of the XRT channels, so it is quite possible that the degradation of the XRT channels needs to be revisited, especially considering the frequent bakeouts.
As we said, the main contributions to the XRT bands, in relatively quiescent active regions, are from \ion{Fe}{xvii} lines. These lines are observable by EIS if there is some microflaring activity, so an estimate of the cross-calibration can be obtained.

The EIS calibration was monitored in-flight until September 2012 \citep{delzanna:2013_eis_calib}. The count rates in the cooler lines in the SW channel did not indicate any significant degradation, while the line ratios between the LW and the SW channels did indicate a strong degradation by about a factor of 2 of the LW channel within the first few years. 

Cross-calibration studies between EIS and the AIA channels  on observations in 2010 have indicated very good agreement \citep{delzanna_etal:2011_aia, delzanna:2013_multithermal}. The cross-calibration of the EIS SW and AIA 193~\AA\ channel is straightforward as both channels observed the same spectral range. 
EIS SW and AIA 193~\AA\ comparisons have been continued 
by one of us (GDZ) and by H. Warren (priv. comm.) and have shown good agreement, within 20\%. 

The cross-calibration with the AIA 94~\AA\ was fine
with observations in 2010 \citep{delzanna:2013_multithermal}. In the Appendix~\ref{2011Oct22}, we present results from an analysis of an EIS observation taken near the time of our observation (on 2011 October 22). In this case, longer exposures were used and a strong signal in the \ion{Fe}{xvii} lines is present. The DEM obtained from the EIS observations predicts AIA 94~\AA\ count rates that are also almost a factor of 2 higher than observed, as in the microflare presented here.

The calibration of the AIA channels is non-trivial. Some of the channels degraded significantly from 2010. Several bakeouts have been carried out to improve sensitivity, and calibration studies are ongoing. The AIA calibration for the first few years has been monitored against the SDO EVE full-Sun spectra. There are no clear indications that the AIA 94~\AA\ degraded significantly (W. Liu, priv.\ comm.).  Relatively good agreement between AIA and EVE in 2010 was found, see e.g. \cite{delzanna_etal:2011_aia}. 

The EVE suffered significant degradation as well, and it was only with the version 5 of its calibration that a technique similar to that one adopted for the EIS calibration (line ratios) was used, improving the resulting irradiances.

\section{Conclusions}
\label{conclusions}

We analysed simultaneous EIS, XRT, and AIA observations of a microflare to obtain the following
new findings: 
a) on the basis of the EIS data an isothermal approximation is reasonable;
b) the temperature obtained from the XRT filter ratio is in very good agreement
with the peak of the DEM obtained from EIS, and only about 4.5~MK; 
c) the small temperature increase (of around 1~MK) of the microflare loop as measured by XRT 
is nearly sufficient to explain the increase in the AIA \ion{Fe}{xviii} line;
d) the microflare loop cools rapidly towards the background temperature, and 
does not appear to cool towards chromospheric temperatures.

We have also shown that:
e) as we found in some  previous cases, the AIA \ion{Fe}{xviii} line is often formed far from the peak abundance of the ion in equilibrium at 7~MK; a reanalysis of SUMER temperatures obtained from 
  \ion{Fe}{xviii} confirms this.
f) The timescale of the event is typical, about 10 minutes. 
In terms of absolute values, we have found significant inconsistencies (factors of about 2) in the cross-calibration of the EIS, XRT, and AIA channels. 
Further work is needed to try and resolve these discrepancies.
However, these inconsistencies do not affect the main results of this paper.

These results encourage us to use the XRT filter ratios to study the 
temperature evolution of all the microflares we have selected in different active regions, 
which will be discussed in a follow-up paper, where 
we will present further evidence that microflares typically reach 
temperatures between 4 and 8~MK. On this basis, it is clear that detailed analyses of microflares need spectroscopic observations of emission lines formed in the 3--10 MK range. We point out that there is no current or planned mission to cover this important temperature range, with the exception of the MaGIXS sounding rocket flight, planned for August 2019.

Finally, it is interesting to note the differences between the microflare presented here
and other small flares of GOES  B and C-class. 
We described `textbook' flares of these two classes in \cite{delzanna_etal:2011_flare} and \cite{petkaki_etal:2012}.
In terms of lightcurves, the larger flares have a steeper increase of the X-ray flux and a more gradual decay, 
which can last much longer. 
Presumably all flares have a similar chromospheric evaporation process occurring at the footpoints, 
as described in detail in \cite{delzanna_etal:2011_flare}. 
Flares of B- and C-class always reach 10-12 MK regardless of the size and energy of the event, 
and their cooling proceeds through all lower temperatures, down to chromospheric temperatures, 
while the loop plasma is draining back to the chromosphere in the same places where it evaporated.
The microflare presented here is reaching at most 5~MK and shows a sudden cooling to 3~MK. We defer the study of this
behaviour with hydrodynamic modelling to a separate paper.

\begin{acknowledgements}
We would like to thank the anonymous referee for very helpful comments, which were addressed in the final version of this paper.
We acknowledge support  by  STFC (UK) via a consolidated grant to the solar/atomic physics group at DAMTP, University of Cambridge. \\

Data supplied are courtesy of the  SDO/AIA consortium.  Hinode is a Japanese
mission developed and launched by ISAS/JAXA, with NAOJ as domestic partner and
NASA and STFC (UK) as international partners. It is operated by these agencies in
co-operation with ESA and NSC (Norway). 
 CHIANTI is a collaborative project involving the University of Cambridge (UK),
George Mason University, the University of Michigan and Goddard Space Flight Centre (USA). 
We found Helioviewer, the European Hinode science data centre, 
and the Lockheed Martin
Solar and Astrophysics Laboratory (LMSAL) AIA data centre very useful.

\end{acknowledgements}


\bibliographystyle{aa}

\bibliography{draft1}  

\begin{thebibliography}{53}
\expandafter\ifx\csname natexlab\endcsname\relax\def\natexlab#1{#1}\fi

\bibitem[{{Asplund} {et~al.}(2009){Asplund}, {Grevesse}, {Sauval}, \&
  {Scott}}]{asplund_etal:2009}
{Asplund}, M., {Grevesse}, N., {Sauval}, A.~J., \& {Scott}, P. 2009, \araa, 47,
  481

\bibitem[{{Brosius} {et~al.}(2014){Brosius}, {Daw}, \&
  {Rabin}}]{brosius_etal:2014}
{Brosius}, J.~W., {Daw}, A.~N., \& {Rabin}, D.~M. 2014, \apj, 790, 112

\bibitem[{{Cargill}(2014)}]{cargill:2014}
{Cargill}, P.~J. 2014, \apj, 784, 49

\bibitem[{{Del Zanna}(1999)}]{delzanna_thesis99}
{Del Zanna}, G. 1999, PhD thesis, Univ.\ of Central Lancashire, UK

\bibitem[{Del~Zanna(2008)}]{delzanna:08_bflare}
Del~Zanna, G. 2008, \aap, 481, L69

\bibitem[{{Del Zanna}(2012)}]{delzanna:12_sxr1}
{Del Zanna}, G. 2012, \aap, 546, A97

\bibitem[{{Del Zanna}(2013{\natexlab{a}})}]{delzanna:2013_eis_calib}
{Del Zanna}, G. 2013{\natexlab{a}}, \aap, 555, A47

\bibitem[{{Del Zanna}(2013{\natexlab{b}})}]{delzanna:2013_multithermal}
{Del Zanna}, G. 2013{\natexlab{b}}, \aap, 558, A73

\bibitem[{{Del Zanna} {et~al.}(2015{\natexlab{a}}){Del Zanna}, {Dere}, {Young},
  {Landi}, \& {Mason}}]{delzanna_etal:2015_chianti_v8}
{Del Zanna}, G., {Dere}, K.~P., {Young}, P.~R., {Landi}, E., \& {Mason}, H.~E.
  2015{\natexlab{a}}, \aap, 582, A56

\bibitem[{{Del Zanna} \& {Ishikawa}(2009)}]{delzanna_ishikawa}
{Del Zanna}, G. \& {Ishikawa}, Y. 2009, \aap, 508, 1517

\bibitem[{{Del Zanna} \& {Mason}(2003)}]{delzanna_mason:2003}
{Del Zanna}, G. \& {Mason}, H.~E. 2003, \aap, 406, 1089

\bibitem[{{Del Zanna} \& {Mason}(2014)}]{delzanna_mason:2014}
{Del Zanna}, G. \& {Mason}, H.~E. 2014, \aap, 565, A14

\bibitem[{{Del Zanna} \& {Mason}(2018)}]{delzanna_mason:2018}
{Del Zanna}, G. \& {Mason}, H.~E. 2018, Living Reviews in Solar Physics, 15

\bibitem[{{Del Zanna} {et~al.}(2011{\natexlab{a}}){Del Zanna}, {Mitra-Kraev},
  {Bradshaw}, {Mason}, \& {Asai}}]{delzanna_etal:2011_flare}
{Del Zanna}, G., {Mitra-Kraev}, U., {Bradshaw}, S.~J., {Mason}, H.~E., \&
  {Asai}, A. 2011{\natexlab{a}}, \aap, 526, A1

\bibitem[{{Del Zanna} {et~al.}(2011{\natexlab{b}}){Del Zanna}, {O'Dwyer}, \&
  {Mason}}]{delzanna_etal:2011_aia}
{Del Zanna}, G., {O'Dwyer}, B., \& {Mason}, H.~E. 2011{\natexlab{b}}, \aap,
  535, A46

\bibitem[{{Del Zanna} {et~al.}(2012){Del Zanna}, {Storey}, {Badnell}, \&
  {Mason}}]{delzanna_etal:12_fe_10}
{Del Zanna}, G., {Storey}, P.~J., {Badnell}, N.~R., \& {Mason}, H.~E. 2012,
  \aap, 541, A90

\bibitem[{{Del Zanna} {et~al.}(2015{\natexlab{b}}){Del Zanna}, {Tripathi},
  {Mason}, {Subramanian}, \& {O'Dwyer}}]{delzanna_etal:2015_emslope}
{Del Zanna}, G., {Tripathi}, D., {Mason}, H., {Subramanian}, S., \& {O'Dwyer},
  B. 2015{\natexlab{b}}, \aap, 573, A104

\bibitem[{{Feldman} {et~al.}(1996{\natexlab{a}}){Feldman}, {Doschek}, \&
  {Behring}}]{feldman_etal:1996a}
{Feldman}, U., {Doschek}, G.~A., \& {Behring}, W.~E. 1996{\natexlab{a}}, \apj,
  461, 465

\bibitem[{{Feldman} {et~al.}(1996{\natexlab{b}}){Feldman}, {Doschek},
  {Behring}, \& {Phillips}}]{feldman_etal:1996b}
{Feldman}, U., {Doschek}, G.~A., {Behring}, W.~E., \& {Phillips}, K.~J.~H.
  1996{\natexlab{b}}, \apj, 460, 1034

\bibitem[{{Feldman} {et~al.}(1992){Feldman}, {Mandelbaum}, {Seely}, {Doschek},
  \& {Gursky}}]{feldman_etal:1992}
{Feldman}, U., {Mandelbaum}, P., {Seely}, J.~F., {Doschek}, G.~A., \& {Gursky},
  H. 1992, \apjs, 81, 387

\bibitem[{{Hannah} {et~al.}(2008){Hannah}, {Christe}, {Krucker}, {Hurford},
  {Hudson}, \& {Lin}}]{hannah_etal:2008}
{Hannah}, I.~G., {Christe}, S., {Krucker}, S., {et~al.} 2008, \apj, 677, 704

\bibitem[{{Hannah} {et~al.}(2016){Hannah}, {Grefenstette}, {Smith}, {Glesener},
  {Krucker}, {Hudson}, {Madsen}, {Marsh}, {White}, {Caspi}, {Shih}, {Harrison},
  {Stern}, {Boggs}, {Christensen}, {Craig}, {Hailey}, \&
  {Zhang}}]{hannah_etal:2016}
{Hannah}, I.~G., {Grefenstette}, B.~W., {Smith}, D.~M., {et~al.} 2016, \apjl,
  820, L14

\bibitem[{{Ichimoto} {et~al.}(1995){Ichimoto}, {Hara}, {Takeda}, {Kumagai},
  {Sakurai}, {Shimizu}, \& {Hudson}}]{ichimoto_etal:1995}
{Ichimoto}, K., {Hara}, H., {Takeda}, A., {et~al.} 1995, \apj, 445, 978

\bibitem[{{Ishikawa} {et~al.}(2014){Ishikawa}, {Glesener}, {Christe},
  {Ishibashi}, {Brooks}, {Williams}, {Shimojo}, {Sako}, \&
  {Krucker}}]{ishikawa_etal:2014}
{Ishikawa}, S.-n., {Glesener}, L., {Christe}, S., {et~al.} 2014, \pasj, 66, S15

\bibitem[{{Kirichenko} \& {Bogachev}(2017)}]{kirichenko_etal:2017}
{Kirichenko}, A.~S. \& {Bogachev}, S.~A. 2017, \apj, 840, 45

\bibitem[{{Klimchuk}(2006)}]{klimchuk:2006}
{Klimchuk}, J.~A. 2006, \solphys, 234, 41

\bibitem[{{Ko} {et~al.}(2009){Ko}, {Doschek}, {Warren}, \&
  {Young}}]{ko_etal:2008}
{Ko}, Y., {Doschek}, G.~A., {Warren}, H.~P., \& {Young}, P.~R. 2009, \apj, 697,
  1956

\bibitem[{{Kobelski} {et~al.}(2014){Kobelski}, {Saar}, {Weber}, {McKenzie}, \&
  {Reeves}}]{kobelski_etal:2014}
{Kobelski}, A.~R., {Saar}, S.~H., {Weber}, M.~A., {McKenzie}, D.~E., \&
  {Reeves}, K.~K. 2014, \solphys, 289, 2781

\bibitem[{{Mrozek} {et~al.}(2018){Mrozek}, {Gburek}, {Siarkowski}, {Sylwester},
  {Sylwester}, {Kepa}, \& {Gryciuk}}]{mrozek_etal:2018}
{Mrozek}, T., {Gburek}, S., {Siarkowski}, M., {et~al.} 2018, \solphys, 293, 101

\bibitem[{{Narukage} {et~al.}(2011){Narukage}, {Sakao}, {Kano}, {Hara},
  {Shimojo}, {Bando}, {Urayama}, {Deluca}, {Golub}, {Weber}, {Grigis},
  {Cirtain}, \& {Tsuneta}}]{narukage_etal:2011}
{Narukage}, N., {Sakao}, T., {Kano}, R., {et~al.} 2011, \solphys, 269, 169

\bibitem[{{Narukage} {et~al.}(2014){Narukage}, {Sakao}, {Kano}, {Shimojo},
  {Winebarger}, {Weber}, \& {Reeves}}]{narukage_etal:2014}
{Narukage}, N., {Sakao}, T., {Kano}, R., {et~al.} 2014, \solphys, 289, 1029

\bibitem[{{O'Dwyer} {et~al.}(2012){O'Dwyer}, {Del Zanna}, {Badnell}, {Mason},
  \& {Storey}}]{odwyer_etal:12_fe_9}
{O'Dwyer}, B., {Del Zanna}, G., {Badnell}, N.~R., {Mason}, H.~E., \& {Storey},
  P.~J. 2012, \aap, 537, A22

\bibitem[{{O'Dwyer} {et~al.}(2014){O'Dwyer}, {Del Zanna}, \&
  {Mason}}]{odwyer_etal:2014}
{O'Dwyer}, B., {Del Zanna}, G., \& {Mason}, H.~E. 2014, \aap, 561, A20

\bibitem[{{Parenti} {et~al.}(2017){Parenti}, {del Zanna}, {Petralia}, {Reale},
  {Teriaca}, {Testa}, \& {Mason}}]{parenti_etal:2017}
{Parenti}, S., {del Zanna}, G., {Petralia}, A., {et~al.} 2017, ArXiv e-prints

\bibitem[{{Parker}(1988)}]{parker:1988}
{Parker}, E.~N. 1988, \apj, 330, 474

\bibitem[{{Peter} {et~al.}(2013){Peter}, {Bingert}, {Klimchuk}, {de Forest},
  {Cirtain}, {Golub}, {Winebarger}, {Kobayashi}, \&
  {Korreck}}]{peter_etal:2013}
{Peter}, H., {Bingert}, S., {Klimchuk}, J.~A., {et~al.} 2013, \aap, 556, A104

\bibitem[{{Petkaki} {et~al.}(2012){Petkaki}, {Del Zanna}, {Mason}, \&
  {Bradshaw}}]{petkaki_etal:2012}
{Petkaki}, P., {Del Zanna}, G., {Mason}, H.~E., \& {Bradshaw}, S.~J. 2012,
  \aap, 547, A25

\bibitem[{Reale(2010)}]{reale:2012_lr}
Reale, F. 2010, Living Rev. Solar Phys., 7

\bibitem[{{Reva} {et~al.}(2015){Reva}, {Shestov}, {Zimovets}, {Bogachev}, \&
  {Kuzin}}]{reva_etal:2015}
{Reva}, A., {Shestov}, S., {Zimovets}, I., {Bogachev}, S., \& {Kuzin}, S. 2015,
  \solphys, 290, 2909

\bibitem[{{Rosner} {et~al.}(1978){Rosner}, {Tucker}, \&
  {Vaiana}}]{rosner_etal:78}
{Rosner}, R., {Tucker}, W.~H., \& {Vaiana}, G.~S. 1978, \apj, 220, 643

\bibitem[{{Saba} \& {Strong}(1991)}]{saba_strong:1991}
{Saba}, J.~L.~R. \& {Strong}, K.~T. 1991, \apj, 375, 789

\bibitem[{{Schmelz} {et~al.}(2015){Schmelz}, {Asgari-Targhi}, {Christian},
  {Dhaliwal}, \& {Pathak}}]{schmelz_etal:2015}
{Schmelz}, J.~T., {Asgari-Targhi}, M., {Christian}, G.~M., {Dhaliwal}, R.~S.,
  \& {Pathak}, S. 2015, \apj, 806, 232

\bibitem[{Schmelz {et~al.}(2012)Schmelz, Reames, von Steiger, \&
  Basu}]{schmelz_etal:2012}
Schmelz, J.~T., Reames, D.~V., von Steiger, R., \& Basu, S. 2012, \apj, 755, 33

\bibitem[{{Shimizu}(1995)}]{Shimizu:1995}
{Shimizu}, T. 1995, \pasj, 47, 251

\bibitem[{{Sterling} {et~al.}(1997){Sterling}, {Hudson}, \&
  {Watanabe}}]{sterling_etal:1997}
{Sterling}, A.~C., {Hudson}, H.~S., \& {Watanabe}, T. 1997, \apjl, 479, L149

\bibitem[{{Sylwester} {et~al.}(2011){Sylwester}, {Sylwester}, {Siarkowski},
  {Engell}, \& {Kuzin}}]{sylwester_etal:2011}
{Sylwester}, B., {Sylwester}, J., {Siarkowski}, M., {Engell}, A.~J., \&
  {Kuzin}, S.~V. 2011, Central European Astrophysical Bulletin, 35, 171

\bibitem[{{Teriaca} {et~al.}(2012){Teriaca}, {Warren}, \&
  {Curdt}}]{teriaca_etal:2012}
{Teriaca}, L., {Warren}, H.~P., \& {Curdt}, W. 2012, \apjl, 754, L40

\bibitem[{{van Ballegooijen} {et~al.}(2011){van Ballegooijen}, {Asgari-Targhi},
  {Cranmer}, \& {DeLuca}}]{van_ballegooijen_etal:2011}
{van Ballegooijen}, A.~A., {Asgari-Targhi}, M., {Cranmer}, S.~R., \& {DeLuca},
  E.~E. 2011, \apj, 736, 3

\bibitem[{{Warren} {et~al.}(2011){Warren}, {Brooks}, \&
  {Winebarger}}]{warren_etal:2011}
{Warren}, H.~P., {Brooks}, D.~H., \& {Winebarger}, A.~R. 2011, \apj, 734, 90

\bibitem[{{Warren} {et~al.}(2012){Warren}, {Winebarger}, \&
  {Brooks}}]{warren_etal:2012}
{Warren}, H.~P., {Winebarger}, A.~R., \& {Brooks}, D.~H. 2012, \apj, 759, 141

\bibitem[{{Watanabe} {et~al.}(1995){Watanabe}, {Haka}, {Shimizu}, {Hiei},
  {Bentley}, {Lang}, {Phillips}, {David Pike}, {Fludra}, {Bromage}, \&
  {Mariska}}]{watanabe_etal:1995}
{Watanabe}, T., {Haka}, H., {Shimizu}, T., {et~al.} 1995, \solphys, 157, 169

\bibitem[{{Winebarger} {et~al.}(2012){Winebarger}, {Warren}, {Schmelz},
  {Cirtain}, {Mulu-Moore}, {Golub}, \& {Kobayashi}}]{winebarger_etal:2012}
{Winebarger}, A.~R., {Warren}, H.~P., {Schmelz}, J.~T., {et~al.} 2012, \apjl,
  746, L17

\bibitem[{{Wright} {et~al.}(2017){Wright}, {Hannah}, {Grefenstette},
  {Glesener}, {Krucker}, {Hudson}, {Smith}, {Marsh}, {White}, \&
  {Kuhar}}]{wright_etal:2017}
{Wright}, P.~J., {Hannah}, I.~G., {Grefenstette}, B.~W., {et~al.} 2017, \apj,
  844, 132

\end{thebibliography}

\appendix  

\section{Additional information on the microflare event of 2011-Sept-3}
\label{add_microflare}

\begin{figure}
\centering
    \includegraphics[width=9cm, angle=0]{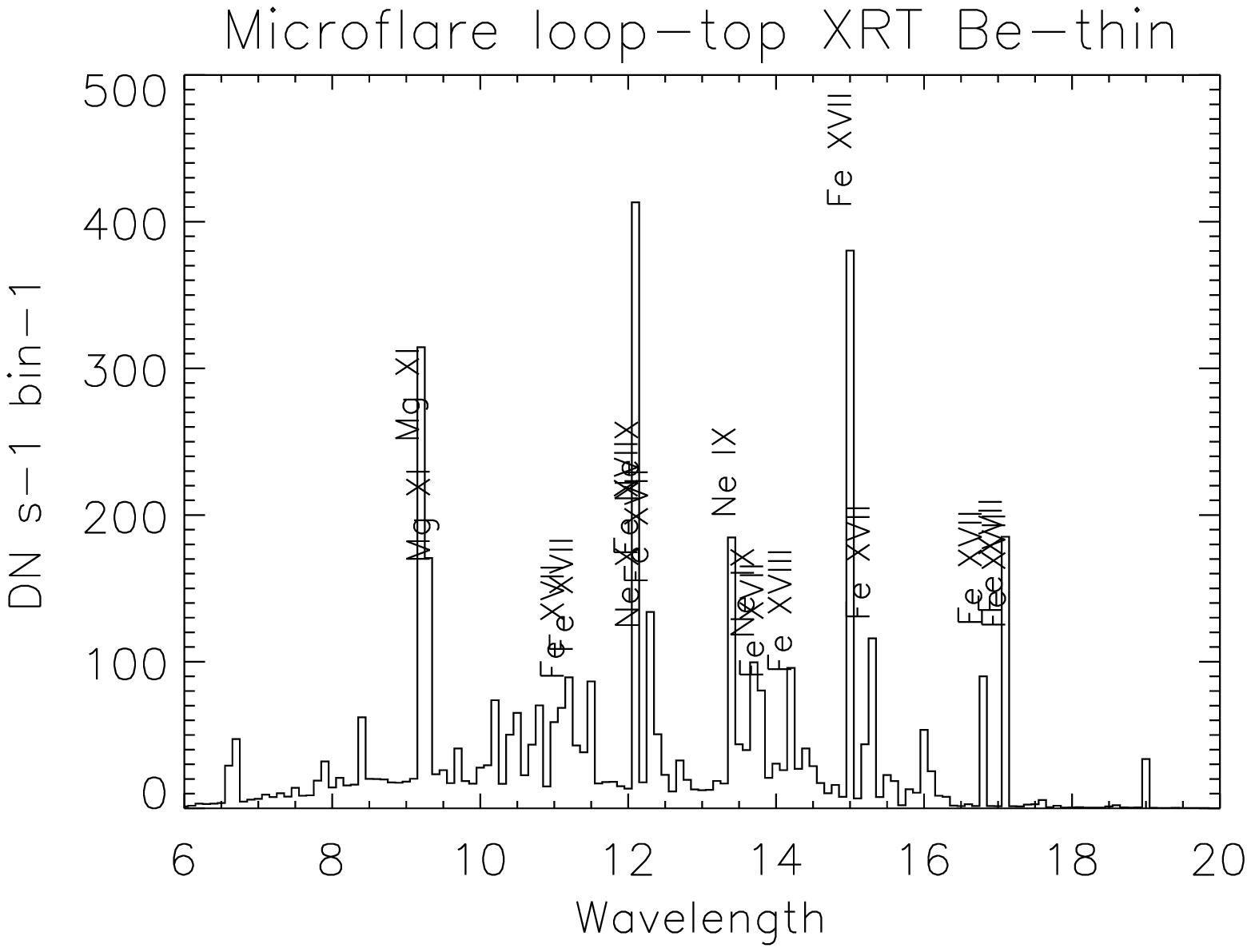}
\centering
    \includegraphics[width=9cm,angle=0]{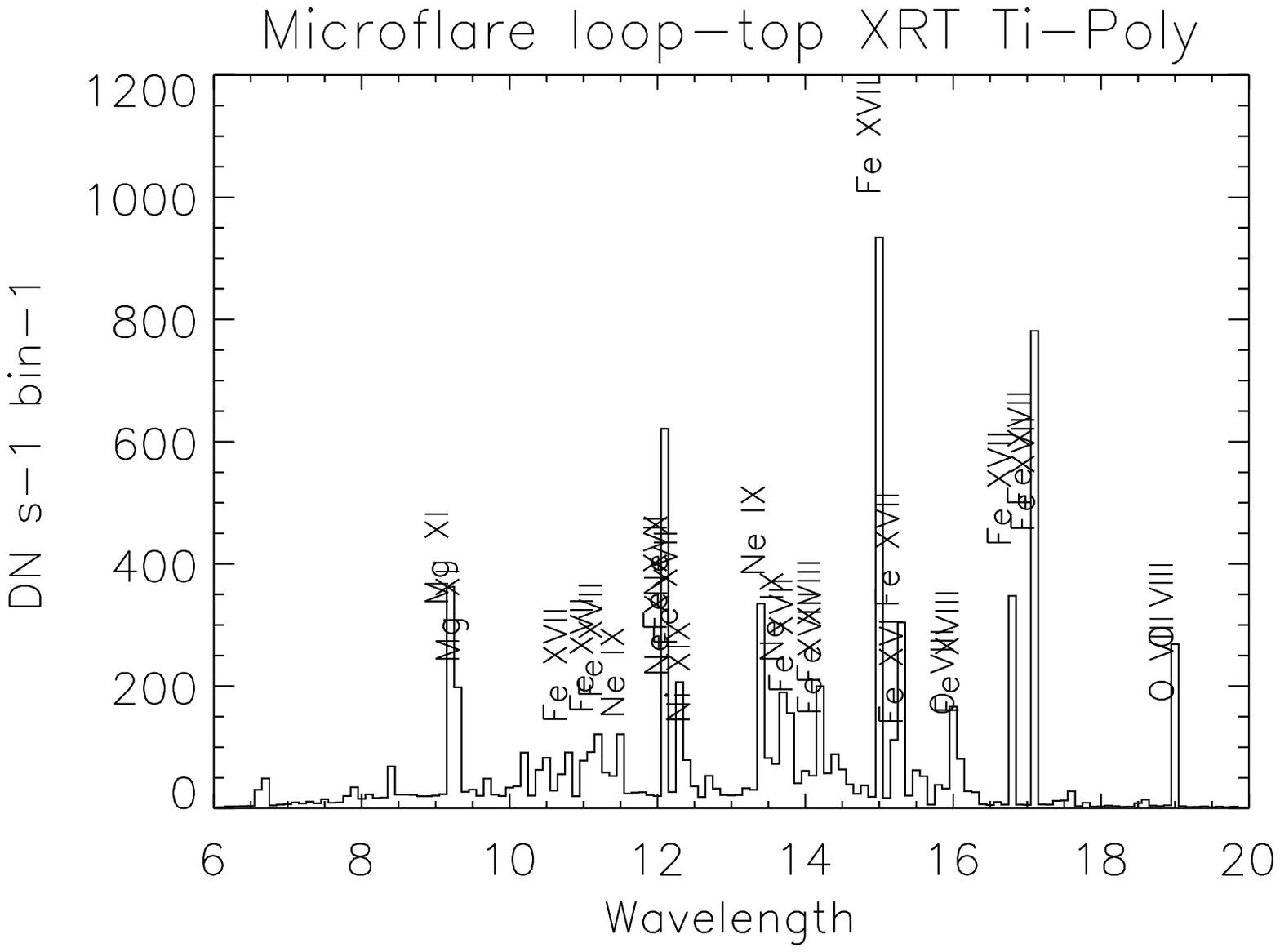}
    \caption{Hinode XRT predicted count rates in the microflare looptop region for the Be-thin and Ti-Poly filters.}
    \label{fig:xrt_sp_microflare}
\end{figure}
Figure~\ref{fig:xrt_sp_microflare} shows the predicted spectra for the Be-thin (top) and the Ti-Poly (bottom) Hinode XRT filters, respectively, in the microflare looptop region, using temperature and emission measure obtained from Hinode EIS.

\begin{figure*}
    \centering
    \includegraphics[width=12cm,angle=90]{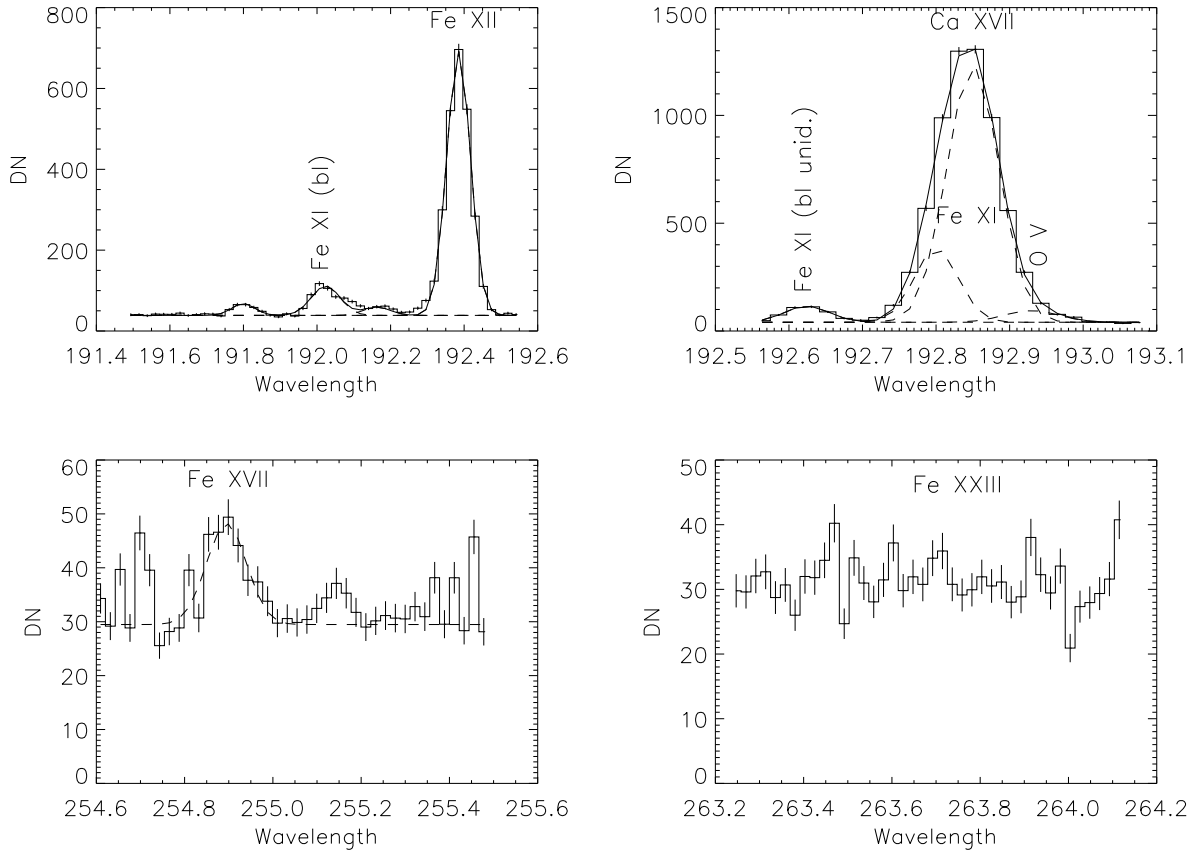}
    \caption{A selection of EIS windows in the microflare looptop region. The spectra are in data numbers (DN) as a function of wavelength (\AA).}
    \label{fig:spectra_microflare}
\end{figure*}
Figure~\ref{fig:spectra_microflare} displays a selection of Hinode EIS lines from the mictroflare looptop region: the well-resolved \ion{Fe}{xi} at 192.0~\AA\ and \ion{Fe}{xii} at 192.4~\AA\ with photon-counts in the hundreds (top left), the just-resolved \ion{Fe}{xvii} at 254.9~\AA\ with a base-to-peak photon-count increase of around 20 (bottom left), an the wavelength region from 263.2$-$264.2~\AA, where the hot \ion{Fe}{xxiii} line would be, which is not detected within the random count fluctuations of around 10~counts (bottom right). The top right image shows the fit to the \ion{Ca}{xvii} line (192.7$-$193.0~\AA), which is blended with \ion{Fe}{xi} as well as \ion{O}{v}, where the dashed lines indicate the individual ions, and the solid smooth line the total of the individual contributions added up.

\section{An EIS vs.\ AIA 94~\AA\ cross-calibration study - 2011 October 22}
\label{2011Oct22}

Considering the discrepancies between the estimated and predicted AIA count rates, we have searched the Hinode EIS database to try and find a suitable observation to have another check on this cross-calibration. 
We have found a full-spectral observation on 2011 October 22, with the 2" slit and 60~s exposures, which therefore has a lot more signal than the microflare observation presented here. The raster duration was about 1~hour. The core of an active region was observed during the last 1/2~hour, when several small microflare loops appeared across the entire AR core.
Several short-lived loops are visible in the AIA 94~\AA\ band, and are also clearly visible in several \ion{Fe}{xvii} lines in EIS. No emission in \ion{Fe}{xxiii} or \ion{Fe}{xxiv} was observed. 

\begin{figure}
\centering
    \includegraphics[width=7cm,angle=0]{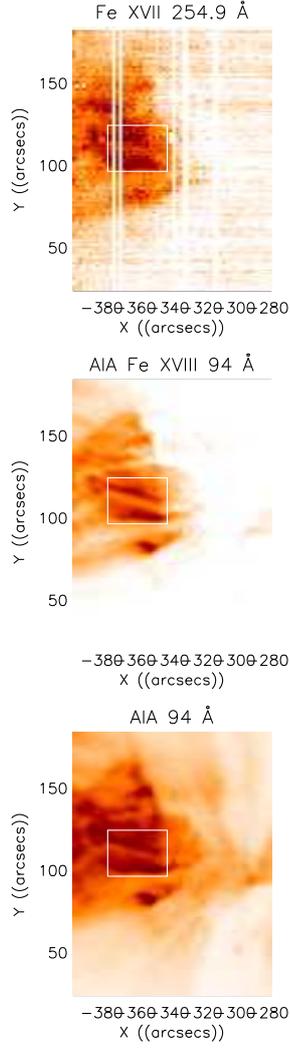}
    \caption{Top: monochromatic image in \ion{Fe}{xvii} of the EIS raster  on 2011 October 22. A microflare region selected for DEM analysis is shown with a box. 
    Middle: estimated \ion{Fe}{xviii} emission in the AIA 94\AA\ band, downgraded to the spatial and temporal resolution of the EIS raster. Bottom: the AIA 94\AA\ band image.}
    \label{fig:aia_eis}
\end{figure}
Figure~\ref{fig:aia_eis} (top) shows a monochromatic image in the strongest \ion{Fe}{xvii} line. The bottom plot in the figure shows a reconstructed AIA 94~\AA\ image, obtained by downgrading the AIA spatial and temporal resolution to that of the EIS observation, following a procedure outlined in, e.g., \cite{delzanna_etal:2011_aia}. The middle plot shows the same AIA image but where the cooler contributions to the band have been removed, following \cite{delzanna:2013_multithermal}. It is clear that an exact agreement between the AIA and EIS images is not present. This is partly due to the difference in the exposure times for each pixel, partly due to the jittering and PSF of the EIS instrument which is not simple to reproduce. 

However, the main features are clearly visible. We took a large area (the box displayed in the Fig.~\ref{fig:aia_eis}) as representative of the ensemble  of microflare loops, and measured the average count rates in the AIA band. 
We obtained 23 DN/s per AIA pixel, and 16 in the estimated \ion{Fe}{xviii} count rates. 

\begin{table}[!htbp]
\caption{Observed  and predicted radiances for the microflares region on 2011 Oct 22
(same layout as Table~\ref{tab:flare}).}
\begin{center}
\footnotesize
\begin{tabular}{@{}llccrlrr@{}}
\hline\hline \noalign{\smallskip}
 $\lambda_{\rm obs}$  & $I_{\rm obs}$ & $T_{\rm max}$ & $T_{\rm eff}$  & $R$ & Ion & $\lambda_{\rm exp}$ &  $r$ \\
   (\AA)             &              &  (log)       & (log)         &     &     &  (\AA)   & \\
\noalign{\smallskip}\hline\noalign{\smallskip}
 171.05 & 3430 &  5.90 &  5.98 &  1.08 &  \ion{Fe}{ix} &  171.073 & 0.98 \\ 
 
 184.52 & 632 &  6.05 &  6.13 &  0.85 &  \ion{Fe}{x} &  184.537 & 0.91 \\ 
 
 192.80 & 263 &  6.13 &  6.18 &  1.06 &  \ion{Fe}{xi} &  192.813 & 0.90 \\ 
 
 180.39 & 2277 &  6.13 &  6.20 &  1.09 &  \ion{Fe}{xi} &  180.401 & 0.97 \\ 
  
 272.00 & 361 &  6.15 &  6.25 &  0.69 &  \ion{Si}{x} &  271.992 & 0.97 \\ 
  
 192.38 & 817 &  6.20 &  6.26 &  1.08 &  \ion{Fe}{xii} &  192.394 & 0.96 \\ 
 
 202.03 & 1703 &  6.25 &  6.32 &  1.08 &  \ion{Fe}{xiii} &  202.044 & 0.97 \\ 
 
 211.30 & 4800 &  6.30 &  6.39 &  1.23 &  \ion{Fe}{xiv} &  211.317 & 0.98 \\ 
  
 284.17 & 28900 &  6.34 &  6.46 &  1.28 &  \ion{Fe}{xv} &  284.163 & 0.99 \\ 
   
 265.02 & 288 &  6.43 &  6.52 &  0.99 &  \ion{Fe}{xvi} &  265.000 & 0.99 \\

 249.19 & 3100 &  6.48 &  6.53 &  0.99 &  \ion{Ni}{xvii} &  249.186 & 0.98 \\ 
 
 193.86 & 347 &  6.57 &  6.55 &  1.28 &  \ion{Ca}{xiv} &  193.866 & 0.98 \\ 
 
 280.17 & 101 &  6.03 &  6.60 &  1.00 &  \ion{Fe}{xvii} &  280.198 & 0.40 \\ 
                                 &   &  &  &  &  \ion{Fe}{xvii} &  280.160 & 0.50 \\ 
 
 200.97 & 263 &  6.64 &  6.60 &  1.06 &  \ion{Ca}{xv} &  200.972 & 0.96 \\ 
 
 269.43 & 49.7 &  6.76 &  6.61 &  1.21 &  \ion{Fe}{xvii} &  269.420 & 0.91 \\ 
  
 254.89 & 127 &  6.74 &  6.62 &  1.19 &  \ion{Fe}{xvii} &  254.885 & 0.99 \\ 
 
 192.89 & 320 &  5.42 &  6.63 &  1.01 &  \ion{Ca}{xvii} &  192.853 & 0.79 \\ 
 
 263.77 & 3 (ul) &  7.16 &  6.91 &  0.90 &  \ion{Fe}{xxiii} &  263.766 & 0.62 \\ 
                                 &   &  &  &  &  \ion{Ni}{xv} &  263.768 & 0.18 \\ 

\noalign{\smallskip}\hline 
\end{tabular}
\end{center}
\normalsize
 \label{tab:flare2} 
\end{table}

\begin{figure}[!htb]
\centering
    \includegraphics[width=8.0cm,angle=0]{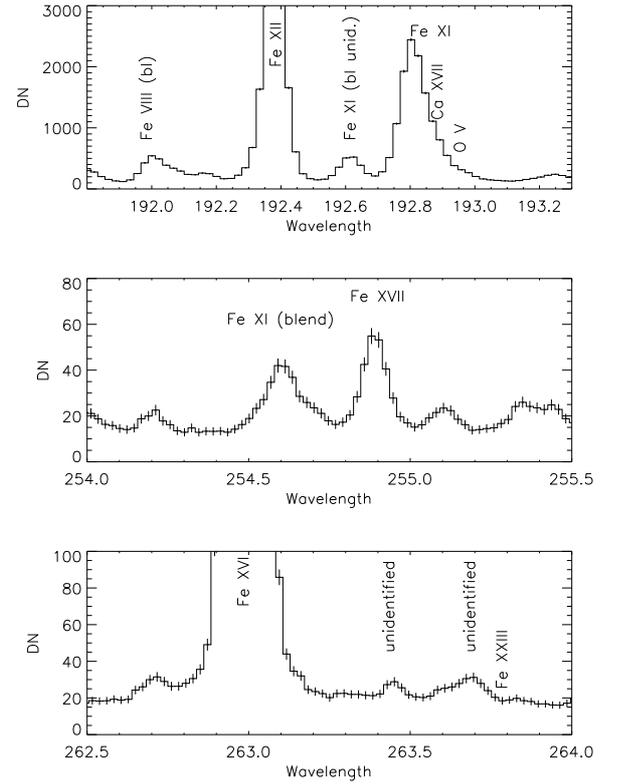}
   \caption{Sample EIS spectral ranges of the microflare region on 2011 October 22.}
    \label{fig:eis_spectra2}
\end{figure}

\begin{figure}[!htb]
\centering
    \includegraphics[width=6.5cm,angle=90]{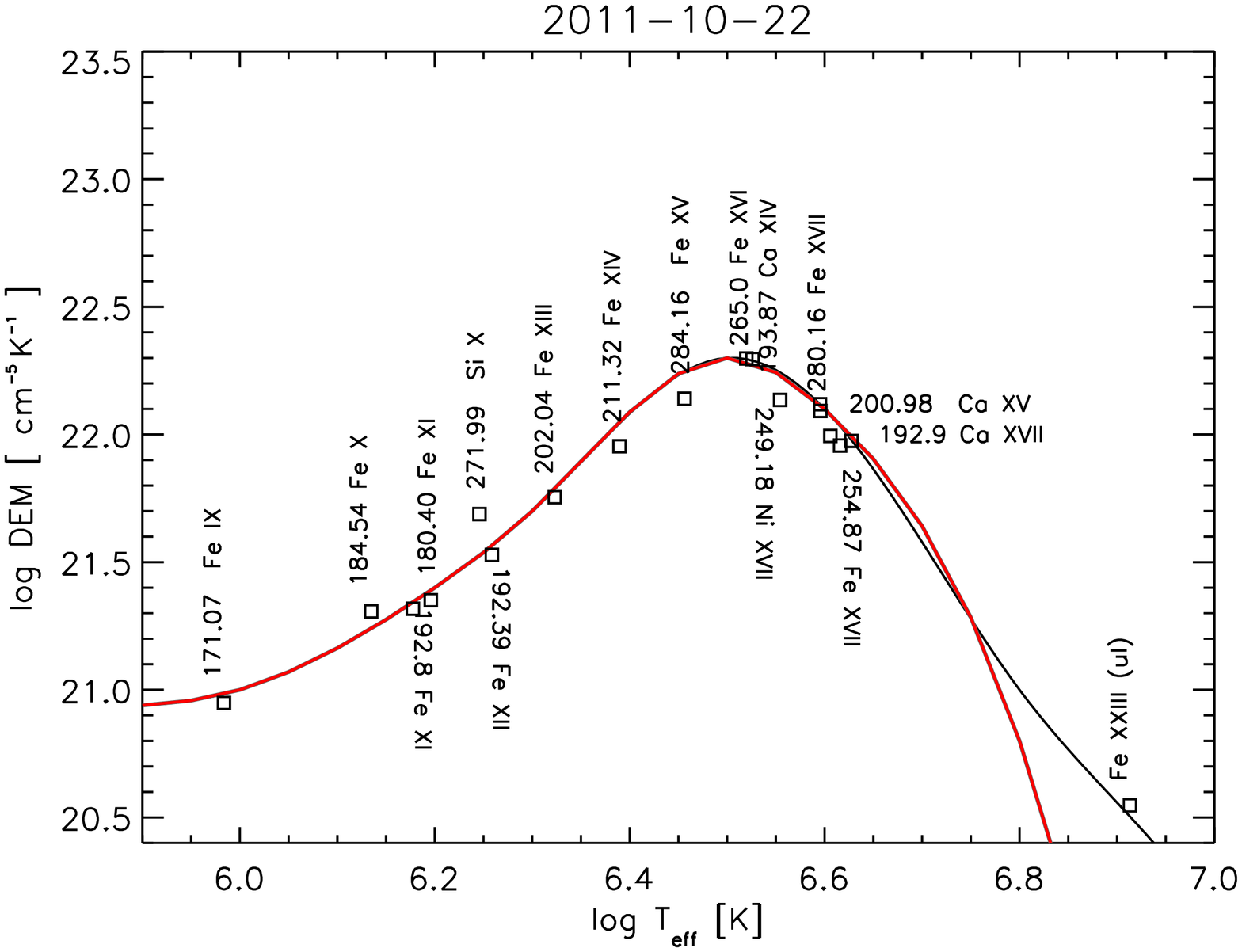}
    \caption{DEM of the microflare region on 2011 Oct 22. The black line was derived using an upper limit for the \ion{Fe}{xxiii} count rates, while the red line had this limit set to 0.}
    \label{fig:aia_eis_dem}
\end{figure}

We averaged the EIS spectra in the area, measured the 
line intensities and performed a DEM analysis. A sample of a few spectral regions is shown in Fig.~\ref{fig:eis_spectra2}. 
The resulting DEM is shown in Fig.~\ref{fig:aia_eis_dem}, and the observed vs. predicted intensities in  Table~\ref{tab:flare2}. 
We found an excellent agreement between observed and predicted intensities in all the main \ion{Fe}{xvii} lines, identified and discussed in \cite{delzanna_ishikawa}.
Excellent agreement between the \ion{Fe}{xvii} and \ion{Ca}{xvii} intensities is present. 
As most of the emission in the area comes from the diffuse AR core emission, we used `coronal abundances', although the predicted AIA count rates are independent of the choice, as all the main lines contributing to the band are from iron.

For the \ion{Fe}{xxiii}, we adopted a similar upper limit as in the other event. Note that the peak ion abundances of \ion{Ca}{xvii} and \ion{Fe}{xvii} are log $T$[K]=6.75, i.e.\ quite close to the peak ion abundance of \ion{Fe}{xviii} ($T$[K]=6.85). As the \ion{Fe}{xviii} ion abundance has a significant tail towards lower temperatures, it turns out that a significant portion of the \ion{Fe}{xviii} emission comes from the same temperature plasma emitting \ion{Fe}{xvii} and \ion{Ca}{xvii}. 
As these ions are also sensitive to emission around 
log $T$[K]=6.85, the slope of the DEM is well constrained at such a temperature. In other words, the estimated \ion{Fe}{xviii} emission is relatively well constrained by the emission measures from \ion{Fe}{xvii} and \ion{Ca}{xvii}. The \ion{Fe}{xxiii} provides a further upper limit of around 10~MK.
With the DEM shown in Fig.~\ref{fig:aia_eis_dem}, we estimate for the AIA \ion{Fe}{xviii} a count rate of 29 DN/s. The \ion{Fe}{xiv} and \ion{Fe}{x} contribute about 4 DN/s and the rest of the lines and continuum another 4. 
Decreasing the DEM in the poorly constrained higher temperatures, as shown in Fig.~\ref{fig:aia_eis_dem} by the red line, only decreases the predicted \ion{Fe}{xviii} signal to 26 DN/s.

This observation from 2011-10-02, which has a much higher signal and more spectral lines than the one from 2011-09-03, confirms the discrepancy between the estimated observed AIA \ion{Fe}{xviii} count rates and the predicted ones from the EIS DEM modelling, with values here of 16~DN/s/pix (observed) versus 29~DN/s/pix (predicted).

\end{document}